\documentclass[useAMS,usenatbib,fleqn]{mn2e}
\usepackage{amsmath}
\usepackage{amssymb}
\usepackage{graphicx}
\newcommand{\g}{g_{\mu\nu}}
\newcommand{\G}{G^\mu_\nu}
\newcommand{\T}{T^{\mu}_{\nu}}

\begin{document}
\title[Cosmological perturbation framework]{A sub-horizon framework for probing the relationship between the cosmological matter distribution and metric perturbations}
\author[Mustafa A. Amin, Robert V. Wagoner and Roger D. Blandford] {Mustafa A. Amin\thanks{E-mail: mamin@stanford.edu},  Robert V. Wagoner and Roger D. Blandford\\ Dept. of Physics and Kavli Institute for Particle Astrophysics and Cosmology, Stanford University, Stanford, CA 94305-4060}

\date{Accepted 2008 May 15. Received 2008 April 9; in original form 2007 August 14}

\pagerange{\pageref{131}--\pageref{142}} \pubyear{2008}

\maketitle
\label{131}

\begin{abstract}
The relationship between the metric and nonrelativistic matter distribution depends on the theory of gravity and additional fields, hence providing a possible way of distinguishing competing theories. With the assumption that the geometry and kinematics of the homogeneous Universe have been measured to sufficient accuracy, we present a procedure for understanding and testing the relationship between the cosmological matter distribution and metric perturbations (along with  their respective evolution) using the ratio of the physical size of the perturbation  to the size of the horizon as our small expansion parameter. We expand around Newtonian gravity on linear, subhorizon scales with coefficient functions in front of the expansion parameter. Our framework relies on an ansatz which ensures that (i) the Poisson equation is recovered on small scales (ii) the metric variables (and any additional fields) are generated and supported by   the nonrelativistic matter overdensity. The scales for which our framework is intended are small enough so that cosmic variance does not significantly limit the accuracy of the measurements and large enough to avoid complications due to nonlinear effects and baryon cooling. From a theoretical perspective, the coefficient functions provide a general framework for contrasting the consequences of $\Lambda$CDM (cosmological constant + cold dark matter) and its alternatives.
We calculate the coefficient functions for general relativity (GR) with a cosmological constant and dark matter, GR with dark matter and quintessence, scalar-tensor theories (STT), $f(R)$ gravity and braneworld (DGP) models. We identify a possibly unique signature of braneworld models. For observers, constraining the coefficient functions provides a streamlined approach for testing gravity in a scale dependent manner. We briefly discuss the observations best suited for an application of our framework. 
\end{abstract}

\begin{keywords}
large scale structure of Universe, dark matter, gravitational lensing.
\end{keywords}

\begin{section}{Introduction}

A successful model of the Universe must include a background geometry,  an inventory of its contents, a kinematical description of its expansion and a dynamical explanation of how its constituents interact, drive the expansion and develop structure. Recent observations (for example \citealt{Riess:1998cb, Perlmutter:1998np, Freedman:2000cf, Allen:2002sr, Tegmark:2003ud, Riess:2004nr, Cole:2005sx, Eisenstein:2005su, Astier:2005qq, Spergel:2006hy, Percival:2007yw, 2008MNRAS.383..879A, 2008MNRAS.387.1179M, Komatsu:2008hk} and references therein) have led to a  ``Flat $\Lambda$CDM cosmology" (henceforth F$\Lambda$CDM),  dominated by dark energy (cosmological constant $\Lambda$) and matter (predominately dark and initially cold) and the observed expansion rate and growth of structure agree with the predictions of this model at the ten percent level. Future observations should be capable of testing this model at the one percent level. If they verify its predictions, they will affirm a remarkable, simple description of the Universe, implicit in the earliest relativistic investigations of Einstein, Friedmann and Lema\^itre, analogous to the affirmation of general relativity (GR) that took place twenty years ago (for example \citealt{lrr-2001-4}). If F$\Lambda$CDM passes this test, then the challenge will be to account for this outcome in terms of physical processes operating at earlier epochs; if it fails, then we shall either have learned something important about gravitational physics or described a new, dominant component of the Universe. Many alternatives, with and without GR, to F$\Lambda$CDM  have been proposed. At this stage, none of them stands out. There is therefore a need to provide a framework for describing future observations and theoretical investigations in general terms which will facilitate a distinction between F$\Lambda$CDM and its alternatives. The provision of one such framework is the goal of this paper.

Further observational progress is anticipated over the coming decade.  The analysis of Planck observations \citep{Planck:2006uk}  
of the microwave background, coupled with local measurements of the contemporary Hubble parameter, $H_0$, should result in an extremely accurate description of the physical conditions and the statistical properties of the density fluctuation spectrum at the epoch of recombination when the Universe had a scale factor $a\equiv (1+z)^{-1}\sim10^{-3}$ relative to today. Combining the calculated physical sizes of the acoustic peaks in the background radiation spectrum with the Hubble constant and the Copernican Principle leads to a measurement of spatial curvature, which is already known to contribute to the kinematics at a level of less than a few percent \citep{Komatsu:2008hk}. We shall adopt a value of zero for illustration purposes. Essentially kinematic measurements, for example, those involving Type Ia supernova explosions, baryonic acoustic oscillations (BAO) and baryonic gas fractions in clusters should provide a record of the comoving distance, $d(a)=\int cdt/a$, from which the evolution of the Hubble parameter $H(a)=d\ln a/dt$ and the acceleration parameter $q(a)=d\ln (Ha)/d\ln a$ can be inferred{\footnote {Our acceleration parameter differs from the conventional deceleration parameter by a minus sign.}}. For the rest of the paper we shall assume that these evolutions have been measured to a sufficient accuracy. Note that we are using $a$ instead of the cosmic time $t$ as the time coordinate as this relates directly to the observable photon frequency shift. For recent constraints on the expansion history, see for example \citep{Rapetti:2006fv} and references therein. 

Given an understanding of the geometry and kinematics, the task is then to see if the dynamical evolution of the Universe is  consistent with GR or mandates an alternative theory. Now, GR provides a relationship between the spacetime geometry on a cosmological scale measured by the Einstein tensor   ${\bf G}[\g]$ and the total Energy-Momentum Tensor (EMT) of its contents $\bf T$,
$ {\bf G}[\g]=8\pi G{\bf T}.$ The discovery  that $ {\bf G}[\g]\ne8\pi G{\bf T}[\textrm{``obs''}]$ where ${\bf T}[\textrm{``obs''}]$ includes known forms of matter such as electromagnetic radiation, baryons etc. has led to the addition of dark matter and dark energy contributions to the EMT. Dark matter candidates include Weakly Interacting Massive Particles and axions which would presumably behave gravitationally like baryonic matter. However other possibilities exist which might behave differently such as massive neutrinos (as a subdominant component).
Dark energy is most simply characterized as a temporaily and spatially constant vacuum energy field with zero enthalpy (see \citealt{Carroll:2000fy} for a review). However, it could also have quite different dynamical properties and might include contributions from additional scalar \citep{Ratra:1987rm}, vector \citep{ArmendarizPicon:2004pm} or tensor  fields with possible interactions between each other \citep{Farrar:2003uw} and with known forms of matter. Historically, the first representation of dark energy was Einstein's cosmological constant, which was seen as an augmentation to $\bf G$, not $\bf T$  (see for example \citealt{Carroll:2003wy}). This original proposal has also been generalized in many ways so that
$ {\bf G}[\g]+{\bf F}[\g,{\bf\varphi}]=8\pi G {\bf T}[\textrm{``obs''}]$, 
where ${\bf F}[\g,{\bf\varphi}]$ depends on the metric and more generally some additional gravitational fields, ${\bf\varphi}$. For example $\varphi$ could be the additional gravitational scalar field in Scalar-Tensor Theories (STT) (see for example \citealt{Santiago:1998ae,Perrotta:1999am}). Nature could of course be unkind and we might have
\begin{equation}
\label{eq:FullField}
{\bf G}[\g]+{\bf F}[\g,{\bf\varphi}]=8\pi G{\bf T}[\textrm{``obs''}]+8\pi G {\bf T}[\textrm{``dark''}].
\end{equation}
Considerable effort has been made in constructing models that fall into the above mentioned categories and more recently in finding ways to distinguish between them (for example see \citealt{Lue:2003ky, Ishak:2005zs,2007soch.conf....9B, 2007PhRvL..99n1302Z, Huterer:2006mv}). 

Now, modifying the physics beyond GR with cold dark matter and $\Lambda$ can have three quite separate manifestations. Firstly it can lead to a change in expansion of the universe, secondly, it can influence the growth of structure and the metric and thirdly, it can confront local tests of the theory of gravity. The approach that we follow is to assume that the theory is constrained by the first and third manifestations and that it is the growth of structure that is providing the test. This oversimplifies the data analysis but does lead to a transparent and simple approach. One important consequence of adopting local gravitational tests is that photons and baryons, at least, will follow geodesics and that the unperturbed photons will be subject to cosmological redshifting of their frequencies, $\nu\propto a^{-1}$. This simplifies the interpretation of observational data.

Our procedure is to adopt a general form for the metric of a linearly perturbed homogeneous and isotropic universe
which introduces two potentials $\Phi({\bf{x}},a)$ and $\Psi({\bf{x}},a)$ (scalar metric perturbations in the Newtonian gauge), where ${\bf{x}}$ denotes the three spatial coordinates.  We also introduce an associated fractional density perturbation $\delta_m({\bf{x}},a)$ in nonrelativistic matter and relate it to  the potentials. We assume that there is a dominant nonbaryonic contribution to the clustering of nonrelativistic matter. In practice, it is easier to work with Fourier modes and this allows us to focus attention on the range of length scales that are most relevant observationally: sufficiently smaller than the horizon so that our expansion is valid and  we can observe enough independent volumes within our current horizon allowing for a high precision measurement despite ``cosmic variance", and yet large enough that nonlinear effects and baryonic cooling are not a factor. Within this range of length scales, we adopt the following ansatz  regarding the relationship between linearized metric and density perturbations, written as an expansion in powers of $(aH/k)$, where $k$ is the magnitude of the comoving wavevector ${\bf{k}}$
\begin{equation}
\begin{aligned}
&\Phi({\bf k},a)=-\frac{4\pi G \rho_m}{H^2}\left(\frac{a H}{k}\right)^{\!\!2}\delta_m({\bf k},a)B\left[\frac{a H}{k},a\right]\\
&\Psi({\bf k},a)=-\frac{4\pi G \rho_m}{H^2}\left(\frac{a H}{k}\right)^{\!\!2}\delta_m({\bf k},a)\Gamma\left[\frac{a H}{k},a\right]\\
&\delta_m({\bf k},a)=\delta_{mi}({\bf k})\Delta\left[\frac{a H}{k},a\right]
\label{ansatz}
\end{aligned}
\end{equation}
where
\begin{equation}
\begin{aligned}
\nonumber
B\left[\frac{a H}{k},a\right]=\left[\beta_0(a)+\beta_1(a)\left(\frac{a H}{k}\right)+\beta_2(a)\left(\frac{a H}{k}\right)^{\!\!2}+\hdots\right]\\
\Gamma\left[\frac{a H}{k},a\right]=
\left[\gamma_0(a)+\gamma_1(a)\left(\frac{a H}{k}\right)+\gamma_2(a)\left(\frac{a H}{k}\right)^{\!\!2}+\hdots\right]\\
\Delta\left[\frac{a H}{k},a\right]=\left[\delta_0(a)+\delta_1(a)\left(\frac{a H}{k}\right)+\delta_2(a)\left(\frac{a H}{k}\right)^{\!\!2}+\hdots\right]
\end{aligned}
\end{equation}
We have set the speed of light $c=1$. The background mass density $\rho_m\propto a^{-3}$. $\delta_{mi}({\bf k})$ is determined from initial conditions which can in principle be taken close to the surface of last scattering, $a_i\sim 10^{-3}$, as long as the modes are sufficiently sub-horizon. Note that often, we are only interested in the scale dependence of the growth of the perturbations in the matter distribution and the metric at linear, subhorizon scales. Such measurements require  taking ratios of the matter distribution or the metric at different redshifts, whereby, the initial conditions cancel out in the final expressions. The coefficient functions $\{\beta_n,\gamma_n,\delta_n\}$ with $n=0,1,2$ are arbitrary functions of the scale factor. The leading terms in the expansion agree with Poisson's equation on small scales, while subsequent terms allow for a scale-dependent departure as we move towards larger scales. This approach introduces a length-scale dependence to the perturbations through an expansion in  powers of $(GM/d_p)^{1/2}\sim d_P/d_H$,  where $M\sim \rho_m d_p^3$ is the total mass enclosed within the physical size $d_p$ and  $d_H\equiv1/H$ is the Hubble horizon. In Fourier space, with $d_p\sim a/k$, we get $(GM/d_p)^{1/2}\sim aH/k$.

The theories that we discuss below introduce different corrections (different coefficient functions  $\{\beta_n,\gamma_n,\delta_n\}$) and these differences are measurable{\footnote{As we shall see at the end of Section 2 and in Section 3, the coefficients $\beta_2,\gamma_2,\delta_2\ne0$ even in GR with non-relativistic matter and represent ``post-Newtonian'' corrections. Also note that $\delta_m$ characterizes the fractional matter density perturbation in the Newtonian gauge, which is related to the often used comoving density perturbation $\Delta_m$ through $\Delta_m=\delta_m+3a\partial_a(\delta_m-3\Psi)(aH/k)^2$. A combination of the $00$ and $0i$ Einstein equations yields $(k/aH)^2\Psi=-(4\pi G\rho_m/H^2)\Delta_m$.}}
. 
From an observer's perspective, constraining the coefficient functions with measurents of $\Phi, \Psi$ and $\delta_m$ provides a streamlined approach to characterizing gravity on cosmological scales in a scale dependent manner. On the other hand, from a theorist's perspective, substituting the ansatz into the field equations for a given theory allows for a (mostly straightforward) calculation of the coefficient functions. The coefficient functions provide a means of comparing the consequences of different theories. We shall discuss our assumptions, limitations and our ansatz in detail in the next section. 

This is certainly not the first time that an attempt at constructing and applying such a framework has been made. The Parametrized Post Newtonian formalism  (see \citealt{lrr-2001-4} and references therein) has been a powerful framework for understanding and constraining gravity on solar system (and other isolated system)  scales. Our aim is to construct a similar framework for cosmological scales. Recently a few attempts have been made in this direction. However most of these are either concerned with the expansion history alone, deal with specific aspects of departures from GR such as effective gravitational constant on small scales \citep{2007PhRvD..76b3514T}, growth of perturbations on small scales \citep{2007APh....28..481L}, the gravitational slip \citep{Caldwell:2007cw}, or deal with superhorizon scales \citep{Bertschinger:2006aw}.  The authors in \citep{2008JCAP...04..013A} take into account growth of structure, anisotropic stress and the modification to the Poisson equation and parametrize departures from Einstein's gravity with a growth index and two functions of the scale factor which are relevant for weak lensing surveys. However, they do not consider scale dependent departures. Another popular phenomenological approach for characterizing the effects of the unknown physics (additional fields, their interactions, or modified gravitational laws) is to define an effective fluid energy momentum tensor for everything other than the standard model matter, effectively move $\bf{F}$ in equation (\ref{eq:FullField}) to the right hand side and define 
${\bf{T}}_{eff}=-(8\pi G)^{-1}{\bf{F}}[\varphi,g_{\mu\nu}]+{\bf{T}}[\textrm{``dark"}]$. This effective energy momentum tensor is then parametrized in terms of the equation of state, sound speed, anisotropic stress, etc. \citep{Hu:1998tj, Bashinsky:2007yc}. This approach, however, seems to put an unnecessary restriction of a fluid interpretation which might be misleading, especially when the effective dark energy is due to modified gravity or extra dimensions. We are unaware of a systematic approach undertaken where the framework includes a scale {\it dependent} departure in the relationship between the matter distribution and the metric perturbations along with their respective evolution on cosmological scales up to post-Newtonian order. \footnote{We note that during the final stages of preparation of this this paper we became aware of a scale dependent framework for modified gravity that includes super and sub-horizon scales \citep{2007PhRvD..76j4043H}. After submission of this manuscript, the following were posted on arXiv.org which are relevant to this work. \citep{Jain:2007yk} provide an analysis of the observational tests for modified gravity;  \citep{2007arXiv0712.1162H} use evolution of galaxy bias to constrain scale dependent departures from GR; whereas \citep{Bertschinger:2008zb} build on \citep{Bertschinger:2006aw}  to include sub-horizon scales; \citep{2008arXiv0801.2433H} extend \citep{2007PhRvD..76j4043H} to include multiple fluids and curvature relevant for cosmic microwave background calculations and constraints; whereas \citep{Caldwell:2007cw} discuss the effects of gravitational slip on the CMB, growth of structure, and lensing observations.}

The rest of the paper is organized as follows. Section 2 discusses our assumptions and the particular form of the ansatz in detail. In Section 3  we apply our framework to GR, STT, quintessence, $f(R)$ models \citep{Carroll:2000fy} and DGP gravity \citep{Dvali:2000hr}. In particular, we calculate the coefficient functions in these theories and comment on our ansatz in the context of these theories. Section 4 is devoted to how our framework might be employed by observers. We briefly discuss the observations that could be used to constrain the different coefficient functions. Section 5 presents a short summary and future directions for extending the framework. 

\end{section}

 \begin{section}{Our ansatz and associated assumptions}
 
With an eye towards observations in the next decade, we assume that the geometry (spatial curvature) and kinematics (expansion history) of the Universe have been measured to a percent level accuracy. What remains to be understood and measured accurately (at the few percent level) is the relationship between the metric fluctuations and the nonrelativistic matter distribution along with their respective evolution on linear, subhorizon scales. This relationship will depend on the theory of gravity or the presence of yet unknown components, thus providing a test for distinguishing different theories. To explore this relationship in an (almost) model independent way, we provide an ansatz, equation (\ref{ansatz}), relating the scalar metric perturbations (in Newtonian gauge) and the nonrelativistic matter overdensity in Fourier space. In this section we discuss the particular form of the ansatz and the underlying assumptions in detail. We introduce our notation and conventions followed by some physical arguments regarding our choice of the particular form of the ansatz. We end with a discussion of the range of scales for which our ansatz is expected to be useful.

We focus on a perturbed FRW universe (spatially flat) with scalar metric fluctuations in the Newtonian gauge \citep{Bardeen:1980kt}. In this gauge the metric takes the following form $(c=1)$
\begin{equation}
ds^2=-[1+2\Phi({\bf x},t)]dt^2+a^2(t)[1-2\Psi({\bf x},t)]d{\bf x}\cdot d{\bf x}\nonumber
\end{equation}
 Here the metric perturbations $|\Phi({\bf x},t)|,|\Psi({\bf x},t)|\ll1$. We choose to work in the Newtonian gauge because $\Phi({\bf x},t)$ is the generalization of the Newtonian gravitational potential and the potentials $\Phi({\bf x},t)$ and $\Psi({\bf x},t)$  are gauge invariant Bardeen variables when we specialize to the Newtonian gauge. The energy density perturbation $\delta_m({\bf x},t)$ is also gauge invariant, corresponding to the energy density perturbation on the zero shear spatial hypersurface which is closest to Newtonian time slicing [see equation (3.14) in \citep{Bardeen:1980kt}].
In what follows, we use the scale factor $a$ as the independent variable instead of cosmic time $t$ with $a(\textrm{today})=1$. With this change of variables, the metric takes the form
\begin{equation}
ds^2=-[1+2\Phi({\bf x},a)](a H)^{\!-2}da^2+a^2[1-2\Psi({\bf x},a)]d{\bf x}\cdot d{\bf x}\nonumber
\end{equation}
We shall work primarily in Fourier space and use the convention $f({\bf x},a)=(2\pi)^{-3}\int d^3{\bf k}f({\bf k},a)e^{i{\bf k}\cdot {\bf x}}$.
To avoid unnecessary clutter we write the Fourier transform of the metric perturbations $\Phi({\bf k},a)e^{i{\bf k}\cdot {\bf x}}$ as $\Phi$. The same is true for $\Psi$ and $\delta_m$.  The background quantities depend on $a$. We shall often suppress this dependence; for example by $H$ we mean $H(a)$.

We have assumed spatial flatness as expected on the basis of the simplest interpretation of inflation. If the Universe has measureable spatial curvature or large scale deviations from the Robertson-Walker assumptions of homogeneity and isotropy, then the following development must be generalized at the expense of introducing parameters that need fitting. A purely geometrical demonstration of spatial flatness would obviate some of this concern. Such a demonstration is possible, in principle, using two screen gravitational lenses (Blandford 2008, in preparation), though it is not known how practical it will be to implement this demonstration. If we choose to include curvature as an additional parameter, then location of the first acoustic peak in the CMB (and BAO scale) would likely provide the best constraints. 
 
Our ansatz provides a relationship between $\Phi$, $\Psi$ and $\delta_m$ on linear (in $\Phi,\Psi$ and $\delta_m$), subhorizon scales. We now turn to the discussion of some important features of this ansatz. On scales that are much smaller than the size of the horizon, $aH/k\ll1$, the leading term has the form of a linearised Newtonian gravitational field equation. For the purpose of this paper the  Newtonian form of the field equation refers to the the following relation between the time-time metric perturbation $\Phi({\bf x},a)$ and the nonrelativistic matter density contrast $\delta_m({\bf x},a)$, $\nabla^2\Phi({\bf x},a)\propto\delta_m({\bf x},a)$, which in Fourier space becomes
$\Phi\propto(aH/k)^2\delta_m$. Now, in the Newtonian gauge $\Phi({\bf x},a)$ plays the role of the Newtonian potential once the background has been subtracted out. The proportionality allows for a possible temporal variation in the effective Newton's constant which could depend on the cosmological background evolution.

From GR we know that this Newtonian relation starts breaking down as the size of the perturbation becomes comparable to the size of the horizon. In general, different theories of gravity will introduce different scale dependent departures from this equation, changing the metric-matter relationship. Our claim is that for a large class of theories, our ansatz, equation (\ref{ansatz}), captures the scale dependence of the relationship between the nonrelativistic matter distribution and cosmological metric perturbations. 
In particular, our ansatz faithfully reproduces the scale dependence of the metric-matter relationship in the fiducial case of GR with cold dark matter and a cosmological constant. In the presence of additional fields one might expect this relationship to break down; however, this is usually not the case. Suppose that an additional field enters the equations, for example as a source (quintessence), as a time varying gravitational constant (Brans Dicke theory) or indirectly encapsulating the effect of higher dimensions, etc. Perturbations $\delta\varphi$ in such a scalar field $\varphi$ (consider quintessence or scalar-tensor theories) will be involved in the relationship between $\delta_m$ and  $\Phi$. However, from the field equation for $\delta\varphi$, equations (\ref{eq:deltaphiST}) and (\ref{eq:deltaphiQ}), we can see that $\delta\varphi\propto\Phi(aH/k)^2$ for quintessence and $\delta\varphi\propto\Phi$ for scalar-tensor theories when $aH/k\ll1$. Thus, even if additional scalar fields are present, our  ansatz should be a good approximation for the relationship between the matter distribution and the metric at the scales of interest. Note that we have assumed $\Psi=\mathcal{O}[\Phi]$ for this argument. 

Another feature of our ansatz is that $\Phi$ and $\Psi$ are directly proportional to $\delta_m$. This might seem unusual, since it  implies that in the absence of nonrelativistic matter perturbations, there would be no metric perturbations. This is certainly not true in principle if an additional scalar field is present. However observationally, we know that nonrelativistic perturbations are present and they dominate over perturbations in other fields. The following argument provides a more detailed justification. Since on the smallest scales, to lowest order in $(aH/k)$, the potential $\Phi\propto\delta_m(aH/k)^2$, we have $\delta\varphi\propto\delta_m(aH/k)^4$ and  $\delta\varphi\propto\delta_m(aH/k)^2$ in quintessence models and STT respectively. This means that the potentials and pertubations in other scalar fields are supported by the nonrelativistic matter perturbations. We do not expect to see the effects of the initial power spectrum of these additional fields up to the order of the terms considered in our ansatz, with the initial power spectrum of the additional field possibly playing a role in higher order terms. This is one of the reasons for not extending the power series in $aH/k$ beyond the order considered in the ansatz. 

Our ansatz does not capture the matter-metric relationship for all available models in the literature. Consider for example $k$-essence \citep{ArmendarizPicon:2000dh},where the effective ``sound speed" ($c_s$) can be small. This leads to a significant clustering of dark energy on small scales which can be comparable to nonrelativistic matter perturbations. In these scenarios, our ansatz does not provide a good approximation to the full theory. The coefficients $\beta_2,\gamma_2(\propto {c_s}^{\!\!-2})\gg1$ signaling a breakdown in our assumptions. More generally, if a model introduces an additional physical scale within the range of scales of interest, then care needs to be taken in using our ansatz. In the $k$-essence example, this additional scale is the Jean's length for the scalar field fluctuations, whereas  in the case of $f(R)$ models this could the ``Compton wavelength" ($\sim f_{RR}^{-1/2}$) of the effective gravitational scalar degree of freedom (see for example \citealt{Pogosian:2007sw}). In such cases the ansatz might still be applicable in a more limited range of parameters and length scales (see  Section 3.4).

We note that some of the above arguments are made under the assumption that the additional gravitational or nongravitational contribution to the field equations is due to a scalar field (quintessence or scalar tensor theories). As argued above, this leads to only even powers of $aH/k$ in the expansion. Furthermore, $f(R)$ modification of the Einstein-Hilbert action also lead to even powers of $aH/k$. An intriguing case where one can get an odd power of $aH/k$ is in DGP braneworld models. In these extra-dimensional theories, the junction conditions on our 4 dimensional brane gives rise to a scale dependence involving terms linear in $aH/k$. We come back to this in Section 3.5. 

\begin{figure}
    \centering
    \includegraphics[width=3in]{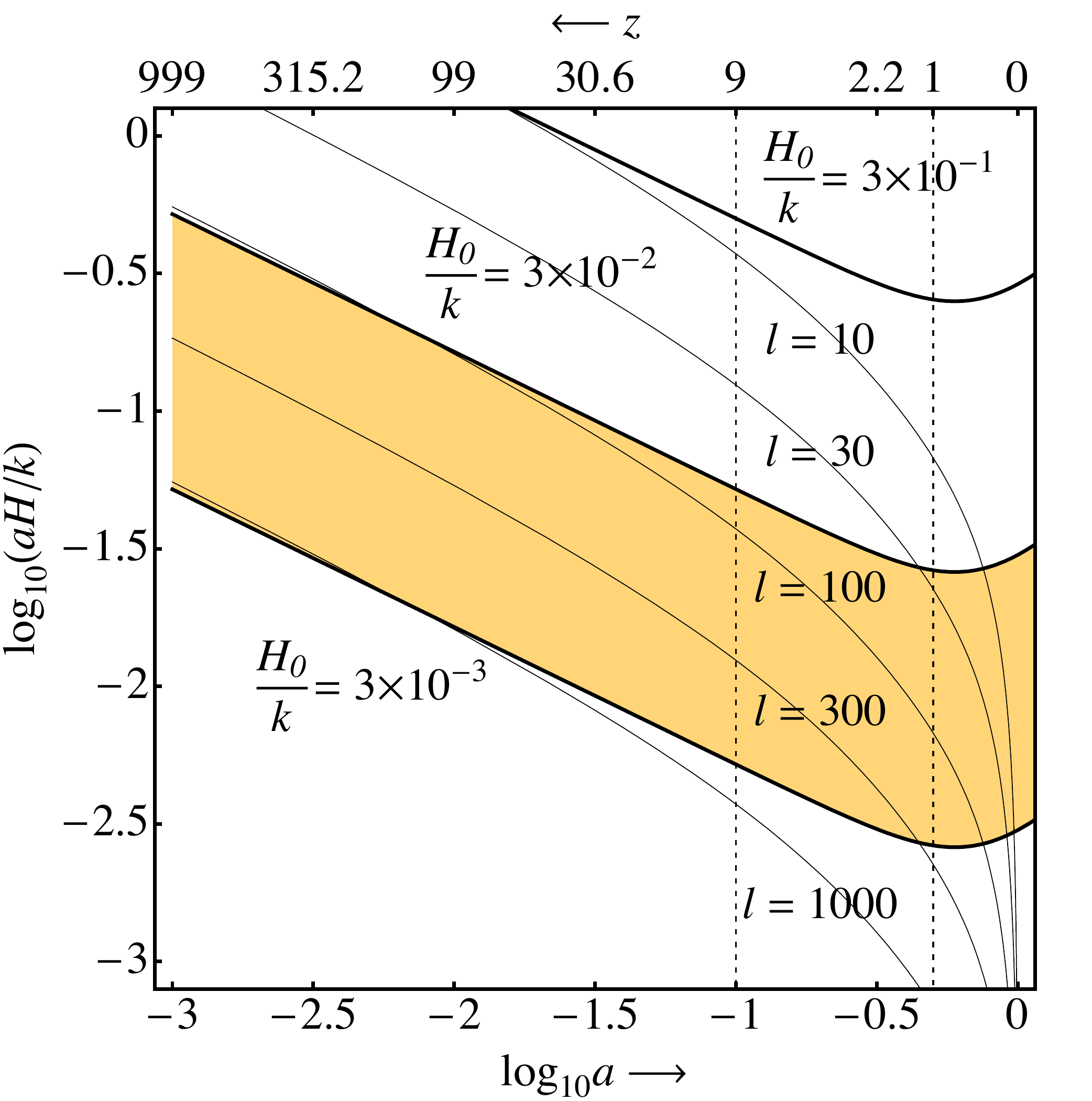} 
    \caption{The ratio of the physical size of the perturbation to the size of the horizon is used as an expansion parameter in our anzatz. We plot this ratio, $(aH/k)$, as a function of $a$ from last scattering to the present for the concordance model (yellow region). The upper and lower bounds of the yellow region are determined by considering scales that are small enough so that cosmic variance does not dominate the errors and at the same time large enough so that nonlinear evolution and baryon cooling are not a significant factor. Most of the observations in the next decade will yield information in the range $10^{-1}\lesssim a\lesssim 1$. If we are interested in observations that only care about a smaller range of the scale factor, then the allowed range of $H_0/k$ increases. We also plot lines of constant multipole $l\sim k d(a)$, which provides a rough estimate of the relationship between $k$ and angular scales at different redshifts.}
    \label{fig: aHoverk}
 \end{figure}

Regarding $\Psi$, we assume that the relationship between $\Psi$ and $\delta_m$ has the same $(aH/k)$ dependence as  $\Phi$ and $\delta_m$ since from GR we expect $\Phi=\Psi$ when no anisotropic stress is present.
The form of $\delta_m({\bf k},a)$ in the ansatz can be motivated from the conservation equation for nonrelativistic matter at first order in $\Phi,\Psi$ and $\delta_m$:
\begin{equation}
\begin{aligned}
\label{eq:cons}
a^2\partial_a^2\delta_m+(2+q)a\partial_a\delta_m=&-\left(\!\frac{k}{aH}\!\right)^{\!\!2}\Phi\\
&+3\left[a^2\partial_a^2\Psi\!+\!(2+q)a\partial_a\Psi\right].
\end{aligned}
\end{equation}
As discussed above at lowest order in $(aH/k)^2$, the metric perturbations $\Phi,\Psi\propto\delta_m(aH/k)^2$, thus the largest term on the RHS of equation (\ref{eq:cons}) is proportional to $\delta_m({\bf k},a)$. At this order we get a homogeneous equation for $\delta_m$ which has a solution of the form $\delta_m({\bf k},a)=\delta_{mi}({\bf k})\delta_0(a)$. This is the usual approximation used when investigating the growth function on small scales. Perturbatively including the next order term on the RHS, we can see that our ansatz captures the general form of the solution to that order. Again, we use this argument as motivation for the form of the ansatz, being aware of the fact that nonrelativistic dark matter is not covariantly conserved in some models. In $\delta_m$, we include both baryonic and nonbaryonic dark matter, with an understanding that baryonic matter contibutes a small fraction to the total. We assume that baryons are covariantly conserved and follow timelike geodesics, serving as test particles whose motion can be used to probe the metric. 

We now turn to a discussion of the range of scales where we expect our procedure to be applicable. Our ansatz uses the ratio of the physical size of the perturbation $d_p(a)$ to the size of the Hubble horizon $d_H(a)\equiv1/H(a)$ as our small (post-Newtonian) expansion parameter. In Fourier space $d_p(a)\sim a/k$  and we need $d_p(a)/d_H(a)\sim{aH}/{k}\ll1$ for the expansion in $aH/k$ to be meaningful. From Figure 1, we can see that for a given $k$, $aH/k$ is a decreasing function of the scale factor (till dark energy domination). So if $aH/k\ll1$ at early times, it will remain so till today. 

We first give a rough upper bound on $H_0/k$. 
In addition to $aH/k\ll1$, from an observational standpoint, the largest scales of interest are the ones where cosmic variance does not significantly limit the precision of our measurements (angular multipoles $l\gg1$). To convert this constraint on $l$ to a constraint on $H_0/ k$, we need a relationship between $k$ and $l$. For large $k$, a perturbation with a given $k$ corresponds roughly to a multipole $l\sim k d(a)=(aH/k)^{-1} aH d(a)$, where $d(a)$ is the co-moving distance. Note that this relationship is scale factor dependent. Let us take $l\sim 30$ as the largest angular scale where cosmic variance does not significantly limit measurement precision. For $0.1\lesssim a\lesssim 0.5$ we get $aH/k\sim aH d(a)/l\lesssim 0.06$ since $aH d(a)\lesssim 3.7$ in this range.  At $a\sim 0.5$, the corresponding comoving wavevector of the perturbation is $k\sim10^{-2}\,h\,\textrm{Mpc}^{-1}$ or equivalently  $H_0/k\sim3\times 10^{-2}$. On the other hand, this same $l$ would correspond to $aH/k>1$ for $a\sim 10^{-3}$. So if we are also interested in the CMB, then $aH/k \lesssim 1$  implies that  $l\sim (aH/k)^{-1} aH d(a)\gtrsim 55$ at $a\sim 10^{-3}$ because $aH d(a)\sim 55$ at last scattering. In summary, for observations at large redshifts, the requirement $aH/k\lesssim1$ provides the upper bound on the scales for which our ansatz can be used wheres $l\gtrsim 30$ does the same as low redshifts. This upper bound  can be relaxed depending on the range of redshift in which the observations are made. 

Now, for the lower bound on $H_0/k$ we get $H_0/k\gtrsim 3\times 10^{-3}$. This corresponds to $k_{nl}\sim10^{-1} \,h\, \textrm{Mpc}^{-1}$  which is at the boundary between linear and nonlinear evolution of $\delta_m$ today. At these scales the linear and nonlinear matter power spectrum differ by a few percent today (and less in the past). Since the scalar metric fluctuations $\mathcal{O}[\Phi({\bf x},a),\Psi({\bf x},a)]\sim 10^{-5}$ on these scales, as indicated by measurements of the cosmic microwave background (CMB), we can linearize the field equations in $\Phi,\Psi$ and $\delta_m$ at these scales. Another reason for this lower bound is that on scales larger than these we do not expect a significant bias between the baryonic and nonbaryonic matter. We can relax the lower bound if the observations are restricted to smaller scale factors since the scale factor dependence of the boundary between linear and nonlinear evolution is given by $k_{nl}(a)\sim 10^{-1}a^{-3/2}\,h\,\textrm{Mpc}^{-1}$. For example if we restrict our selves to $10^{-3}\lesssim a\lesssim 10^{-1}$, then $H_0/k\gtrsim 10^{-4}$.

Figure 1 shows the typical order of magnitude of $aH/k$ for the range  $3\times10^{-3}\lesssim H_0/k \lesssim 3\times10^{-2}$ (filled yellow region). Finally, the range of scale factors we have in mind for our framework is $10^{-1}\lesssim a\lesssim 1$. Gravitational dynamics at late times (large $a$) is particularly interesting due to cosmic acceleration. The next generation of observations including lensing, BAO, cluster counts, galaxy power spectra etc. will be made within this range. 
Although we concentrate on late times, with some care, our framework can be used with CMB observations. 
For example, after including radiation and baryons, using our framework we can calculate the  anisotropies in the CMB if we know the initial conditions for each mode $\it{after}$ it enters the horizon. Once the modes are sufficiently subhorizon, their subsequent evolution can be used to constrain the coefficient functions. Note, that for the mode corresponding to the first acoustic peak ($l\sim 220$), $aH/k\sim 0.3$ at last scattering. This comoving scale (as well as a range of smaller scales) is within the yellow shaded region in Figure 1.

Before we end this section we provide a concrete example of what the coefficient functions look like in a simple case, the Einstein-de Sitter universe:
\begin{equation}
\begin{aligned}
&\beta_0=\gamma_0=1,\\
&\beta_1=\gamma_1=0,\\
&\beta_2=\gamma_2=-3,\\
&\delta_0=a/{a_i},\\
&\delta_1=0,\\
&\delta_2=3(a/a_i)(1-a/a_i).
\end{aligned}
\end{equation}
where $a_i\sim 10^{-2}$. We turn to the  calculation of the coefficient functions in the next section. 

\end{section}
\begin{section}{Application of the framework with examples}

In this section we calculate the coefficient functions for GR with a cosmological constant and nonrelativistic matter, GR with quintessence, scalar-tensor theories, $f(R)$ theories and DGP gravity. In general, the nonrelativistic matter consists of  baryons, massive neutrinos and nonbaryonic dark matter with (possibly) nongravitational interactions between them and other fields.  For simplicity we will ignore massive neutrinos and baryons in this section. Local tests of gravity provide strong constraints on baryons and photons and their interactions. They do not yet provide similar constraints on the interactions of nonbaryonic matter. Hence, nonbaryonic matter need not be covariantly conserved. However in the examples considered, we treat dark matter as a perfect fluid that is covariantly conserved for simplicity. This allows us to use the conservation equation (\ref{eq:cons}), which is sometimes easier to use than a gravitational field equation that would otherwise take its place. 
 
The basic strategy is to substitute our ansatz into the field equations and conservation equations and solve for the coefficient functions. We begin by substituting our ansatz (\ref{ansatz}) into the conservation equation for nonrelativistic perfect fluid dark matter (\ref{eq:cons}), collecting terms with like powers of $(aH/k)$ and setting their coefficient terms equal to zero to obtain 
\begin{equation}
\begin{aligned}
\label{eq:cons_coeff}
&\left[a^2\partial_a^2+(2+q)a\partial_a\right]\delta_0-\frac{4\pi G \rho_m}{H^2}\beta_0\delta_0=0,\\
&\left[a^2\partial_a^2+(2+q)a\partial_a\right][(aH)\delta_1]-\frac{4\pi G \rho_m}{H^2}\beta_0[(aH)\delta_1]\\
=&\frac{4\pi G \rho_m}{H^2}(aH)\beta_1\delta_0,\\
&\left[a^2\partial_a^2+(2+q)a\partial_a\right][(aH)^2\delta_2]-\frac{4\pi G \rho_m}{H^2}\beta_0[(aH)^2\delta_2]\\
=&\frac{4\pi G \rho_m}{H^2}(aH)^2[2\beta_1\delta_1+\beta_2\delta_0-3(a^2\partial_a^2+qa\partial_a-q)(\gamma_0\delta_0)],
\end{aligned}
\end{equation}
where $q(a)$ and $H(a)$ are assumed to be known from the background evolution. The above equations are second order differential equations for $\delta_0,\delta_1$ and $\delta_2$. The equation for $\delta_0$ can be solved once $\beta_0$ is known. $G\beta_0$ is the effective gravitational ``constant". If $\beta_0=1$, the equation for $\delta_0$ is the usual equation for the fractional matter overdensity on linear and  small scales in GR with nonrelativistic matter as the only clustering component. 

We digress a bit to note that for $\bar{\delta}_n\equiv(aH)^n\delta_n$, the differential operator acting on $\bar{\delta}_n$  is $[a^2\partial_a^2+(2+q)a\partial_a-4\pi G\rho_m\beta_0/H^2]$. This feature continues if we were to go to higher order terms as well, hence it might be useful to find a Green's function for this operator.  In general, to solve for $\delta_1$, we need to know $\beta_0,\beta_1$ and $\delta_0$ with two initial conditions. Similarly, to solve for $\delta_2$ we need to know $\beta_0,\gamma_0,\delta_0, \beta_1,\delta_1$ and $\beta_2$ along with two initial conditions. To progress further we turn to specific theories of gravitation. Our aim is to show how to apply the formalism rather than discuss in detail the various models considered. We leave out the detailed steps, which are straightforward but tedious. 

 
\begin{subsection}{General relativity with cold dark matter and the cosmological constant}

We start with the usual Einstein Hilbert action:
\begin{equation}
\begin{aligned}
&S=\frac{1}{16\pi G} \int d^4x \sqrt{-g}\left[R-2\Lambda\right]+ \int d^4x \sqrt{-g}\mathcal{L}_{m},
\end{aligned}
\end{equation}
with $\mathcal{L}_m$, the lagrangian density for perfect fluid cold dark matter
The corresponding field equations are 
\begin{equation}
\G+\Lambda \delta^\mu_\nu=8\pi G {}\T,
\end{equation}
where $\G=R^\mu_\nu-\delta^\mu_\nu R/2$ and $\T$ is the energy-momentum tensor for a pressureless perfect fluid. As usual, we separate the field equations into the background and perturbed parts (first order in $\Phi,\Psi$ and $\delta_m$). Upon substitution of our ansatz into the perturbed field equations  we get the following expressions/equations for the coefficient functions.
\begin{equation}
\begin{aligned}
&\left[a^2\partial_a^2+(2+q)a\partial_a\right]\delta_0-\frac{4\pi G \rho_m}{H^2}\delta_0=0,\\
&\left[a^2\partial_a^2+(2+q)a\partial_a\right][(aH)\delta_1]-\frac{4\pi G \rho_m}{H^2}[(aH)\delta_1]=0,\\
&\left[a^2\partial_a^2+(2+q)a\partial_a\right][(aH)^2\delta_2]-\frac{4\pi G \rho_m}{H^2}[(aH)^2\delta_2]\\
=&-\frac{12\pi G \rho_m}{H^2}(aH)^2\left[a^2\partial_a^2+(q+1)a\partial_a-q\right]\delta_0,\\
&\beta_0=\gamma_0=1,\\
&\beta_1=\gamma_1=0,\\
&\beta_2=\gamma_2=-3\frac{a\partial_a\delta_0}{\delta_0},\\
\end{aligned}
\end{equation}
where we used the $00$ and $i\ne j$ Einstein equations along with the coefficient form of the conservation equations (\ref{eq:cons_coeff}). We need to provide 6 constants of integration for the three second order differential equations. We take these to be 
\begin{equation}
\begin{aligned}
\label{eq:ini}
&\delta_0(a_i)=1,\quad\quad a_i\partial_a\delta_0(a_i)=1,\\
&\delta_1(a_i)=0,\quad\quad a_i\partial_a\delta_1(a_i)=0,\\
&\delta_2(a_i)=0,\quad\quad a_i\partial_a\delta_2(a_i)=-3.
\end{aligned}
\end{equation}
This ensures that $\delta_m({\bf k},a_i)=\delta_{mi}({\bf k})$, thus defining $\delta_{mi}({\bf k})$ in our ansatz (\ref{ansatz}). The derivatives are chosen to agree with the case of pure matter domination at early times ($a_i\sim 10^{-2}$), where the explicit solution takes the form $\delta_0=a/a_i$, $\delta_1=0$ and $\delta_2=3(a/a_i)(1-a/a_i)$ after rejecting the decaying modes. For any model under consideration, we can choose fix initial condition by rejecting the decaying mode. For simplicity, we shall use the above initial conditions for the scalar-tensor as well the braneworld models for which we plot the coefficient functions. For these models the parameters have been chosen so that at $a_i\sim 10^{-2}$, the conservation equations approach those of an Einstein-deSitter universe in GR.
 
The dashed lines in Figures 2 and 3 show these dimensionless coefficient functions for the spatially flat-$\Lambda$CDM with $\Omega_m=8\pi G\rho_{m0}/3H_0^2=0.3$. Since $\beta_0=\gamma_0=1$, there are no corrections to the Newtonian gravitational constant as far as growth of perturbations is concerned on small scales. Since single gradients do not appear in the Einstein equations involving $\delta_m,\Phi$ and $\Psi$ (after eliminating the velocity through the conservation equation), $\beta_1=\gamma_1=0$. The $00$ Einstein equation imposes $\delta_1=0$. The fact that $\beta_2=\gamma_2\ne0$ reflects corrections because of GR to the relationship between matter and metric perturbations, whereas $\beta_2=\gamma_2\ne-3$ reflects the effect of the cosmological constant. $\delta_0$ characterizes the growth of structure on small scales. It deviates from $\delta_0=a/a_i$ because of $\Lambda$. $\delta_2$ reflects the corrections to the growth function as we move to larger scales. Note that $\beta_2$ and $\gamma_2$ and $\delta_2$ are multiplied by $(aH/k)^2$, whose magnitude is shown in Figure 1. The terms $\beta_2(aH/k)^2$, $\gamma_2(aH/k)^2$ and $\delta_2(aH/k)^2$ are much smaller than $\beta_0$, $\gamma_0$ and $\delta_0$, making it difficult to observe their effects unless we investigate large scales. 
 \end{subsection}

\begin{subsection}{Scalar-tensor theory with cold dark matter (matter representation)}

Scalar-tensor theories are popular alternatives to GR. In the matter representation (also called the Jordan frame), the action contains two free functions $f(\varphi)$ and $V(\varphi)$
\begin{equation}
\begin{aligned}
&S=\frac{1}{16\pi G}\int d^4x \sqrt{-g}\left[f(\varphi)R+\mathcal{L}_\varphi\right]+\int d^4x \sqrt{-g}\mathcal{L}_{m}.
\end{aligned}
\end{equation}
Note that we have decided to make $\varphi$ dimensionless since we wish to treat the perturbation in this field $\delta\varphi$ on the same footing as the metric perturbations $\Phi$ and $\Psi$. Also $\mathcal{L}_\varphi=-(\partial \varphi)^2/2-V(\varphi)$ and $\mathcal{L}_m$ does not contain $\varphi$. The field equations for this theory are
\begin{equation}
\begin{aligned}
&G^\mu_\nu+\frac{1}{f}\left[\delta^\mu_\nu\Box-\nabla^\mu \nabla_\nu\right]f\\
=&\frac{8\pi G}{f} T^\mu_\nu+\frac{1}{2f}\left[\partial^\mu\varphi\partial_\nu\varphi-\delta^\mu_\nu\left(\frac{1}{2}\partial^\sigma\varphi\partial_\sigma\varphi+V\right)\right].\\
\end{aligned}
\end{equation}
The field equation for $\varphi$ is 
\begin{equation}
\label{eq:STfield}
\Box \varphi-V_\varphi+f_\varphi R=0,
\end{equation}
where $f_\varphi=\partial_\varphi f$ and $V_\varphi=\partial_\varphi V$. 

These field equations at the background level can be found in the literature (for example see \citealt{Boisseau:2000pr}). Using our ansatz in the perturbed gravitational field equations and the field equations for $\varphi$ at first order in $\Phi,\Psi,\delta_m$ and $\delta\varphi$, collecting terms with like powers of $(aH/k)$, and setting the expression in front of each power of $(aH/k)$ equal to zero, we get the following expressions/equations for the coefficient functions:
\begin{equation}
\begin{aligned}
&\beta_0=\frac{1}{f}\left(\frac{1+4f\alpha^2}{1+3f\alpha^2}\right)\approx \frac{1}{f}+\mathcal{O}[\alpha^2],\\
&\gamma_0=\frac{1}{f}\left(\frac{1+2f\alpha^2}{1+3f\alpha^2}\right)\approx \frac{1}{f}+\mathcal{O}[\alpha^2],\\
&\beta_1=\gamma_1=0,\\
&\beta_2=-\frac{3}{f}\frac{a\partial_a\delta_0}{\delta_0}+\frac{1}{4f^2}(a\partial_a\varphi)^2\\
&\quad\quad+\left[-3(a\partial_a\varphi)\frac{a\partial_a\delta_0}{\delta_0}+\frac{1}{2}(a\partial_a\varphi)^2\frac{\alpha_\varphi}{\alpha}+3(a\partial_a\varphi)+\frac{3V_\varphi}{2H^2}\right]\frac{\alpha}{f}\\
&\quad\quad+\mathcal{O}[\alpha^2],\\
&\gamma_2=-\frac{3}{f}\frac{a\partial_a\delta_0}{\delta_0}+\frac{1}{4f^2}(a\partial_a\varphi)^2\\
&\quad\quad+\left[\left(a\partial_a\varphi\right)\frac{a\partial_a\delta_0}{\delta_0}+\frac{1}{2}(a\partial_a\varphi)^2\frac{\alpha_\varphi}{\alpha}-(a\partial_a\varphi)-\frac{V_\varphi}{2H^2}\right]\frac{\alpha}{f}\\
&\quad\quad+\mathcal{O}[\alpha^2],\\
\end{aligned}
\end{equation}
where $\alpha=f_\varphi/f$ is the coupling function and all the functions depend on the scale factor $a$ . We have calculated the full expressions for $\beta_2$ and $\gamma_2$, which are rather long. The first two terms are listed as a power series in the coupling function $\alpha\ll1$ with $\alpha\sim\alpha_\varphi,\alpha_{\varphi\varphi}...$ . We used the $i\ne j$ equation, $\alpha\delta\varphi=\Psi-\Phi$, to eliminate $\delta\varphi$ from the field equations. The $00$ equation and the field equation for $\delta\varphi$ yield $\beta_n$ and $\gamma_n$ with ($n=0,1,2$). The equations for $\delta_0,\delta_1$ and $\delta_2$ are given by equations (\ref{eq:cons_coeff}) with  $\beta_n$ and $\gamma_n$ ($n=0,1,2$) given above. Again using the initial conditions (\ref{eq:ini}), we can solve for all the coefficient functions once $f(\varphi)$ and $V(\varphi)$ have been provided. Note that the difference $\Phi-\Psi$ depends on $\beta_n-\gamma_n$ ($n=0,2$). This is usually small for $\alpha\ll1$ since $\beta_0-\gamma_0\sim \alpha^2$ and $\beta_2-\gamma_2\sim\alpha$. 
\begin{figure}
    \centering
    \includegraphics[width=3in]{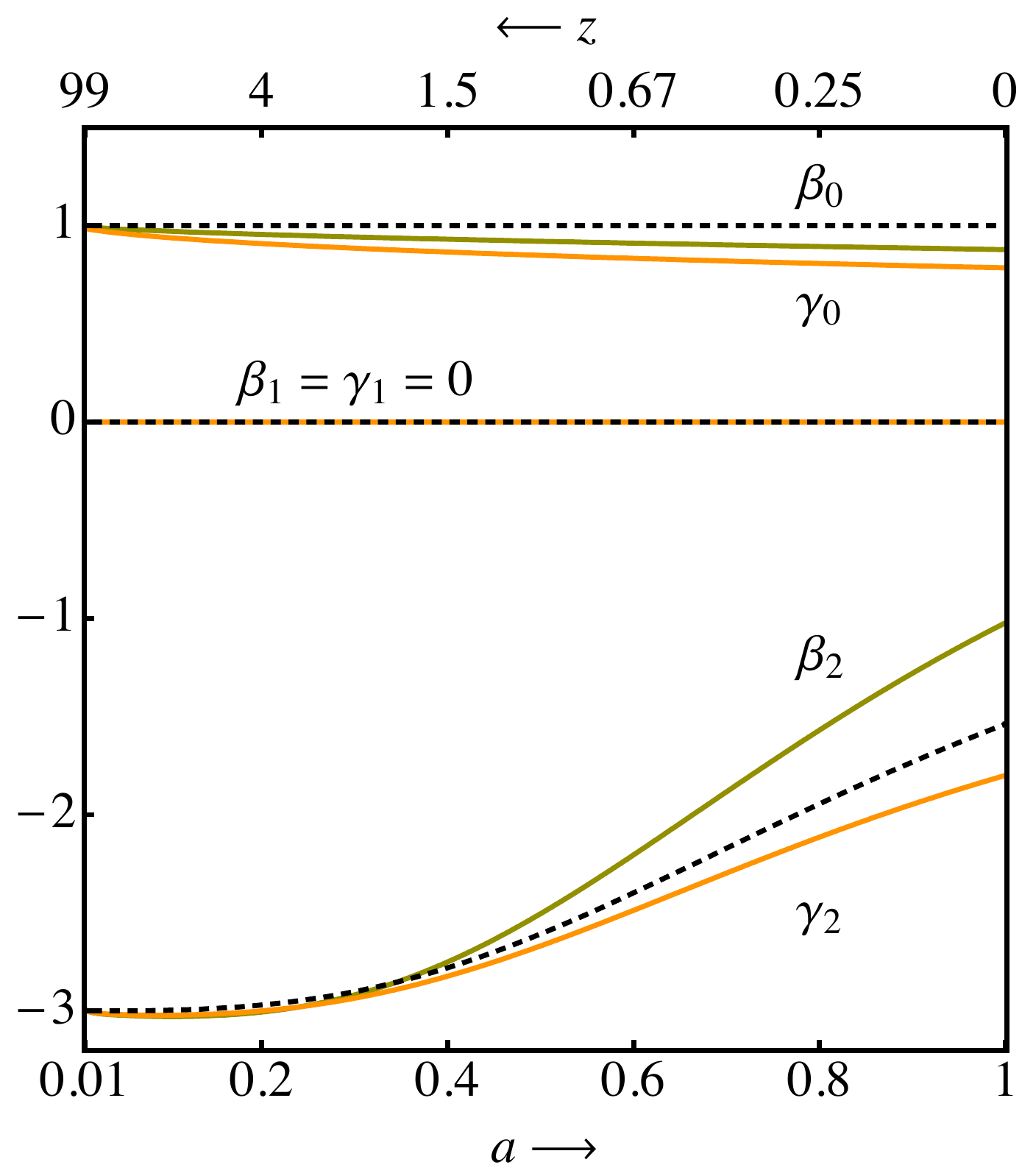} 
    \caption{The dimensionless coefficient functions characterizing the relationship between the metric perturbations and matter distribution are shown above for $F\Lambda$CDM(dashed lines) and the scalar-tensor theory (STT) (solid lines). The STT model is chosen so that its expansion history is consistent with observations. In the case of $\Lambda$CDM $\beta_0=\gamma_0=1$, $\beta_1=\gamma_1=0$ and $\beta_2=\gamma_2$. At early time (matter domination)  $\beta_2=\gamma_2=-3$ with the cosmological constant causing a departure from this value at late times. The variation of $\beta_0$ with the scale factor in the STT can be interpreted as a variation of Newton's constant ``$G\beta_0$'' as far as growth of perturbations is concerned. Also note that for STT, $\beta_0\ne\gamma_0$ and $\beta_2\ne\gamma_2$. For STT, the difference in the coefficient functions is due to $\Phi-\Psi=-\alpha(\varphi)\delta\varphi\ne0$. Note that $\beta_1=\gamma_1=0$ in STT as well as $\Lambda$CDM. We remind the reader that in the ansatz (\ref{ansatz}) the coefficients $\beta_2$ and $\gamma_2$ are multiplied by $(aH/k)^2$, whose magnitude is shown in Figure 1, making them accessible at large scales only.}
    \label{fig: coeff_2}
 \end{figure}
\begin{figure}
    \centering
    \includegraphics[width=3in]{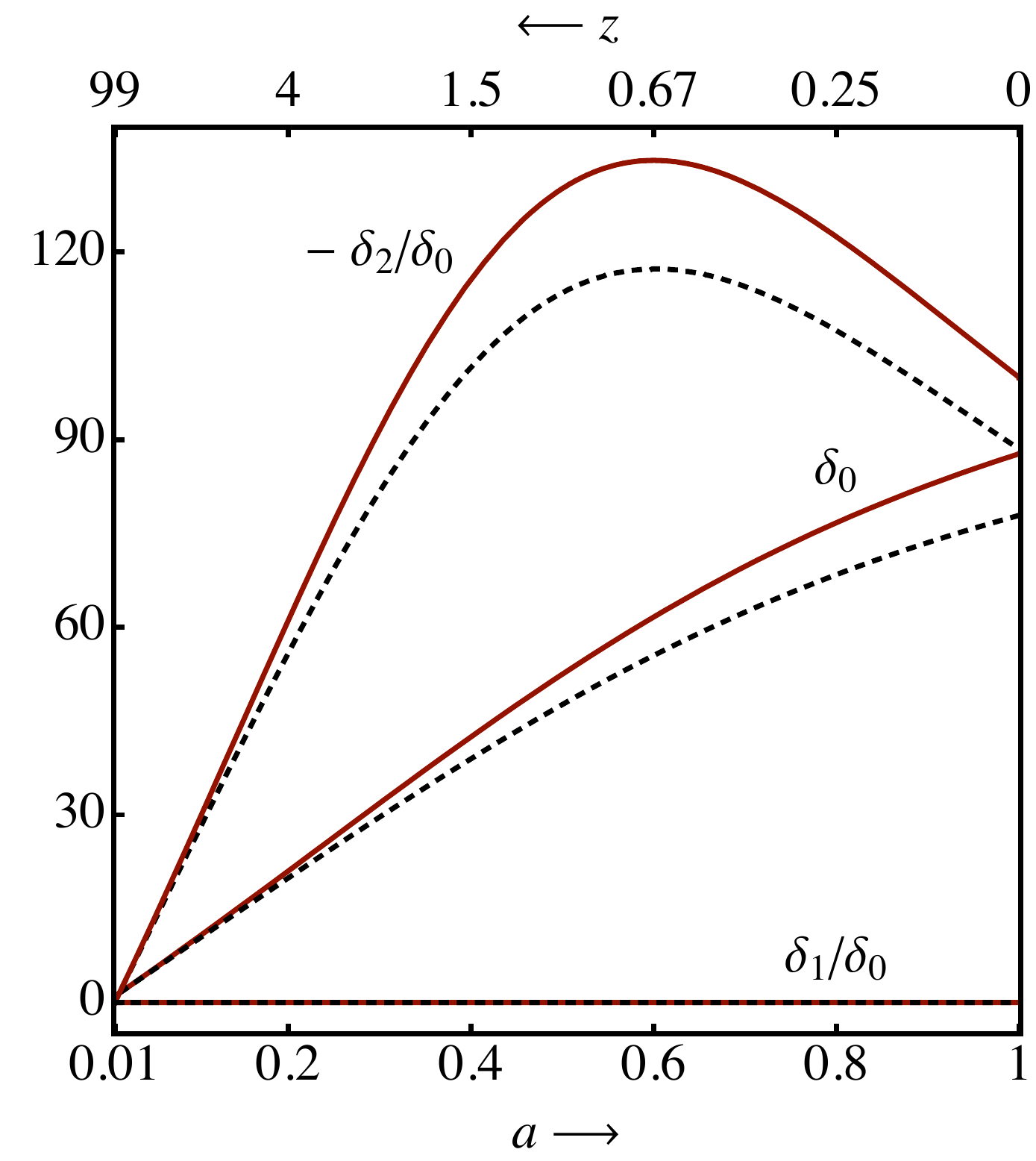} 
    \caption{The dimensionless coefficient functions characterizing growth of structure are show above for $\Lambda$CDM(dashed lines) and the scalar-tensor theory (STT) (solid lines). The STT model is chosen so that its expansion history is consistent with observations. $\delta_0$ is the usual growth function on small scales, whereas $\delta_2$ characterizes the departures as we move to larger scales. For $\Lambda$CDM and STT, $\delta_1=0$. We note that $\delta_2$ is the coefficient of $(aH/k)^2$, which is small withing the scales of interest (see Figure 1). The initial conditions for $\delta_0$ and $\delta_2$ are chosen at $a_i\sim10^{-2}$ and are consistent with growth of structure in a matter dominated era.}
    \label{fig: coeff_3}
 \end{figure}
 
We plot the coefficient functions in Figures 2 and 3. We have chosen $f(\varphi)=1+c_1\varphi^2$ and $V(\varphi)=2\Lambda(1+c_2\varphi^2)$ with $c_1=c_2=0.1$. The initial conditions and $c_1,c_2$ were chosen to ensure that the expansion history remains consistent with observations (consistent with $\Lambda$CDM to within a few percent). The difference between $\beta_n$ and $\gamma_n$ ($n=0,2)$ is due to nonminimal coupling ($\alpha\ne0$). We stress that we have not included baryons in this illustrative calculation. Including baryons would lead to very strong constraints on the function $f(\varphi)$ today from solar system tests \citep{Schimd:2004nq}. For an example of a STT that includes dark matter and baryons with different couplings to gravity see \citep{Bean:2001ys}. 

\end{subsection}
 
\begin{subsection}{General relativity with cold dark matter and quintessence}

GR with quintessence is a special case of the scalar-tensor theories discussed above, with $f(\varphi)=1$. The action and corresponding field equations are
\begin{equation}
\begin{aligned}
&S=\frac{1}{16\pi G} \int d^4x \sqrt{-g}\left[R+\mathcal{L}_{\varphi}\right]+ \int d^4x \sqrt{-g}\mathcal{L}_{m}
\end{aligned}
\end{equation}
\begin{equation}
\begin{aligned}
&\G=8\pi G \T+\frac{1}{2}\left[\partial^\mu\varphi\partial_\nu\varphi-\delta^\mu_\nu\left(\frac{1}{2}\partial^\sigma\varphi\partial_\sigma\varphi+V\right)\right]\\
\end{aligned}
\end{equation}
\begin{equation}
\label{eq:Qfield}
\Box \varphi-V_\varphi=0
\end{equation}
The coefficient functions are given by

\begin{equation}
\begin{aligned}
&\beta_0=\gamma_0=1,\\
&\beta_1=\gamma_1=0,\\
&\beta_2=\gamma_2=-3\frac{a\partial_a\delta_0}{\delta_0}+\frac{1}{4}(a\partial_a\varphi)^2\\
\end{aligned}
\end{equation}
where $(a\partial_a\varphi)^2/4=1-q-4\pi G\rho_m/H^2$. The $i\ne j$ Einstein equation yields $\beta_n=\gamma_n$ ($n=0,1,2$). We used the $0i$ equation to eliminate $\delta\varphi$ from the field equations. As before $\delta_0,\delta_1$ and $\delta_2$ are provided by equation (\ref{eq:cons_coeff}).

We pause to comment on a difference between minimally and nonminimally coupled scalar-tensor theories. Consider the field equation (\ref{eq:STfield}) for $\delta \varphi$: 
\begin{equation}
\begin{aligned}
\label{eq:deltaphiST}
&\left[a^2\partial_a^2+(3+q)a\partial_a\right] \delta \varphi+\left[\left(\!\frac{k}{aH}\!\right)^{\!\!2}+\frac{V_{\varphi\varphi}}{H^2}-6(1+q)f_{\varphi\varphi}\right]\delta\varphi\\
&= (a\partial_a\varphi-6f\alpha)a\partial_a\Phi+3(a\partial_a\varphi-2(4+q)f\alpha)a\partial_a\Psi\\
&\quad-2\left(6f\alpha(1+q)+\frac{V_\varphi}{H^2}\right)\Phi+2f\alpha\left(\!\frac{k}{aH}\!\right)^{\!\!2}(\Phi-2\Psi)
\end{aligned}
\end{equation}
In the minimally coupled case we set $f(\varphi)=1,\alpha(\varphi)=0$ to get
\begin{equation}
\begin{aligned}
\label{eq:deltaphiQ}
&\left[a^2\partial_a^2+(3+q)a\partial_a\right] \delta \varphi+\left[\left(\frac{k}{aH}\right)^{\!\!2}+\frac{V_{\varphi\varphi}}{H^2}\right]\delta\varphi\\
=& (a\partial_a\varphi)a\partial_a\Phi+3(a\partial_a\varphi)a\partial_a\Psi-2\frac{V_\varphi}{H^2}\Phi.
\end{aligned}
\end{equation}
 From the above equations we can see that in the nonminimally coupled case, for $k/aH\gg1$ we have $\delta\varphi\propto\alpha(\varphi)\Phi$ whereas in the minimally coupled case $\delta\varphi\propto\Phi(aH/k)^2$. Along with $\Phi\propto\delta_m(aH/k)^2$, at large $k$ the additional field $\delta\varphi$ follows the same $aH/k$ expansion as the potentials with $\delta_{mi}({\bf k})$ multiplying the expansion. This is one of the arguments we had used in Section 2 to justify the form of our ansatz. We have assumed $\Psi=\mathcal{O}[\Phi]$ in this argument.
\end{subsection}
 
\begin{subsection}{$f(R)$ gravity with cold dark matter}

In recent years modifications of the Einstein-Hilbert action in the form of a function of the Ricci scalar has become a popular alternative to quintessence \citep[see for example][]{Carroll:2003wy, Nojiri:2006ri}. The action and field equations are
\begin{equation}
\begin{aligned}
&S=\frac{1}{16\pi G} \int d^4x \sqrt{-g}\left[R+f(R)\right]+ \int d^4x \sqrt{-g}\mathcal{L}_{m}\,\\
&(1+f_R)\G-\delta^\mu_\nu\frac{f}{2}+[\delta^\mu_\nu\Box-\nabla^{\mu}\nabla_{\nu}]f_R={8\pi G} \T\, .
\end{aligned}
\end{equation}
In the above expressions $f_R=\partial_Rf(R)$. The coefficient functions are
\begin{equation}
\begin{aligned}
&\beta_0=\frac{4}{3(1+f_R)},\\
&\gamma_0=\frac{2}{3(1+f_R)},\\
&\beta_1=\gamma_1=0,\\
&\beta_2=\frac{1}{(1+f_R)}\left[\frac{2}{3}\frac{a^2\partial_a^2\delta_0}{\delta_0}\right.\\
&\quad\quad-\left.\frac{2}{3}\left\{(24B(j+q-2)+2-q\right\}\frac{a\partial_a\delta_0}{\delta_0}\right.
\\&\quad\quad\left.+4B\left\{10-4q+q^2+2j(q-4)-s\right\}\right.\\
&\quad\quad\left.+72B^2(j+q-2)^2-4(j+q-2)a\partial_aB+2q-\frac{1}{9B}\right],\\
&\gamma_2=\frac{1}{(1+f_R)}\left[-\frac{2}{3}\frac{a^2\partial_a^2\delta_0}{\delta_0}\right.\\
&\quad\quad\left.+\frac{2}{3}\left\{(6B(j+q-2)-7-q\right\}\frac{a\partial_a\delta_0}{\delta_0}\right.
\\&\quad\quad\left.-4B\left\{4-q+q^2+j(2q-5)-s\right\}\right.\\
&\quad\quad\left.+4(j+q-2)a\partial_aB-2q+\frac{1}{9B}\right],\\
\end{aligned}
\end{equation}
where $j=dq/d\ln a-(1-2q)q$  and $s= dj/d\ln a-(2-3q)j$ are the scale factor dependent functions, jerk and snap respectively, and $B=H^2f_{RR}/(1+f_R)$\footnote{Our $B=H^2f_{RR}/(1+f_R)$ differs from the definition of $B$ in \citep{Song:2006ej} by a factor of $(q-1)/6(j+q-2)$. Also it is unrelated to the $B$ used in (\ref{ansatz})}.  To obtain $\delta_0,\delta_1$ and $\delta_2$ we use equation (\ref{eq:cons_coeff}). Again, as in the case of GR and scalar-tensor example, the coefficients of the $aH/k$ term in the ansatz vanish. Note that we have assumed $(aH/k)^2B^{-1}\ll1$ in deriving the above expressions, hence it is not appropriate to take the limit $B\rightarrow 0$ after deriving the coefficient functions. Under this assumption, to lowest order in $aH/k$, we get $\Phi=2\Psi$, unlike GR with $\Lambda$CDM where $\Phi=\Psi$. If we take the opposite limit, $(aH/k)^2 B^{-1}\gg1$ the coefficient functions are quite different. In particular, $\beta_0=\gamma_0=(1+f_R)^{-1}$ and we reach the GR limit as we let $f_R\rightarrow 0$. As long as we ensure, a priori, that this transition scale $(H/k_C)\sim B^{1/2}$ (see for example \citealt{Pogosian:2007sw}) is outside the length range of physical scales of interest, we can use our ansatz.  
More details on the dynamics of $f(R)$ theories in the context of structure formation, solar system tests,  etc. can be found in \citep{2007PhRvD..76f3505F,Song:2006ej,Bean:2006up,Chiba:2006jp,Pogosian:2007sw}. Finally we note, that our purpose in discussing $f(R)$ models was to illustrate an application of our framework. These models suffer from a number of problems including fine tuning to match the solar system constraints as well as a rather serious instability,  where the curvature blows up at finite matter densities  \citep{Frolov:2008uf}. 
\end{subsection}

\begin{subsection}{Brane world models: DGP Gravity}

As a final example, we provide the expressions and equations governing the coefficient functions for DGP gravity. In this model, matter is restricted to a four dimensional brane in a five dimensional bulk. In addition to the the Einstein-Hilbert action in the bulk, there is an induced four dimensional term \citep{Dvali:2000hr}. More explicitly, the full 5D action is given by 
\begin{equation}
\begin{aligned} 
&S=\frac{1}{32\pi G r_c} \int d^5 x \sqrt{- g_{(5)}}R_{(5)}\\
&\quad\quad+\frac{1}{16\pi G} \int d^4x \sqrt{-g}R+\int d^4x \sqrt{-g}\mathcal{L}_{m}\,.
\end{aligned}
\end{equation}
In the above action $r_c=G^{(5)}/2G$ where $G^{(5)}$ is 5D gravitational constant. The field equations are given by the Einstein equations in the bulk  ($A,B=0,1,2,3,4$):
\begin{equation}
\begin{aligned} 
&G^A_B=0,
\label{EinsteinBulk}
\end{aligned}
\end{equation}
and the Israel junction conditions on the brane ($\mu,\nu=0,1,2,3$)
\begin{equation}
\begin{aligned} 
&K^\mu_\nu=r_c\left(\mathcal{G}^\mu_\nu-\frac{1}{3}\mathcal{G}\delta^\mu_\nu\right)\,,
\label{Israel}
\end{aligned}
\end{equation}
where $K^\mu_\nu$ is the extrinsic curvature on the brane. In Gaussian normal co-ordinates, the extrinsic curvature is given by the derivative normal to the brane
\begin{equation}
\begin{aligned} 
K^\mu_\nu=\frac{1}{2}\partial_y g_{\mu\nu}\,.
\label{GN}
\end{aligned}
\end{equation}
On the RHS of the junction conditions
\begin{equation}
\begin{aligned} 
\mathcal{G}^\mu_\nu=G^\mu_\nu-8\pi GT^\mu_\nu\,,
\label{RHS}
\end{aligned}
\end{equation}
where $G^\mu_\nu$ and $T^\mu_\nu$ are the 4D Einstein and stress-energy tensors respectively. For this model our coefficient functions are given by (which can be easily determined from the results in \citealt{Deffayet:2002fn})
\begin{equation}
\begin{aligned}
&\beta_0=\frac{4-2Hr_c(2+q)}{3-2Hr_c(2+q)},\\
&\gamma_0=\frac{2-2Hr_c(2+q)}{3-2Hr_c(2+q)},\\
&\beta_1=\frac{12(-1+Hr_c(1+q))^2}{Hr_c(3-2Hr_c(2+q))^2},\\
&\gamma_1=\frac{6(1-2 H r_c)(1-Hr_c(1+q))}{Hr_c(3-2Hr_c(2+q))^2}.
\end{aligned}
\end{equation}
We can solve for $\delta_0$ and $\delta_1$ using (\ref{eq:cons_coeff}). Note that even though $\delta_1(a_i)=a_i\partial_a\delta_1(a_i)=0$, $\delta_1(a)\ne0$ because $\beta_1\ne0$.  We note an important difference between the the DGP braneworld model and the examples considered so far in this paper.  Unlike the previous examples, the coefficients of the odd power of $aH/k$ are non-zero ($\beta_1,\gamma_1,\delta_1\ne 0$). As explained below, the odd power of $aH/k$ arises due to the junction conditions that must be satisfied by metric perturbations at the location of our four dimensional brane in the higher dimensional bulk.  

\begin{figure}
    \centering
    \includegraphics[width=3in]{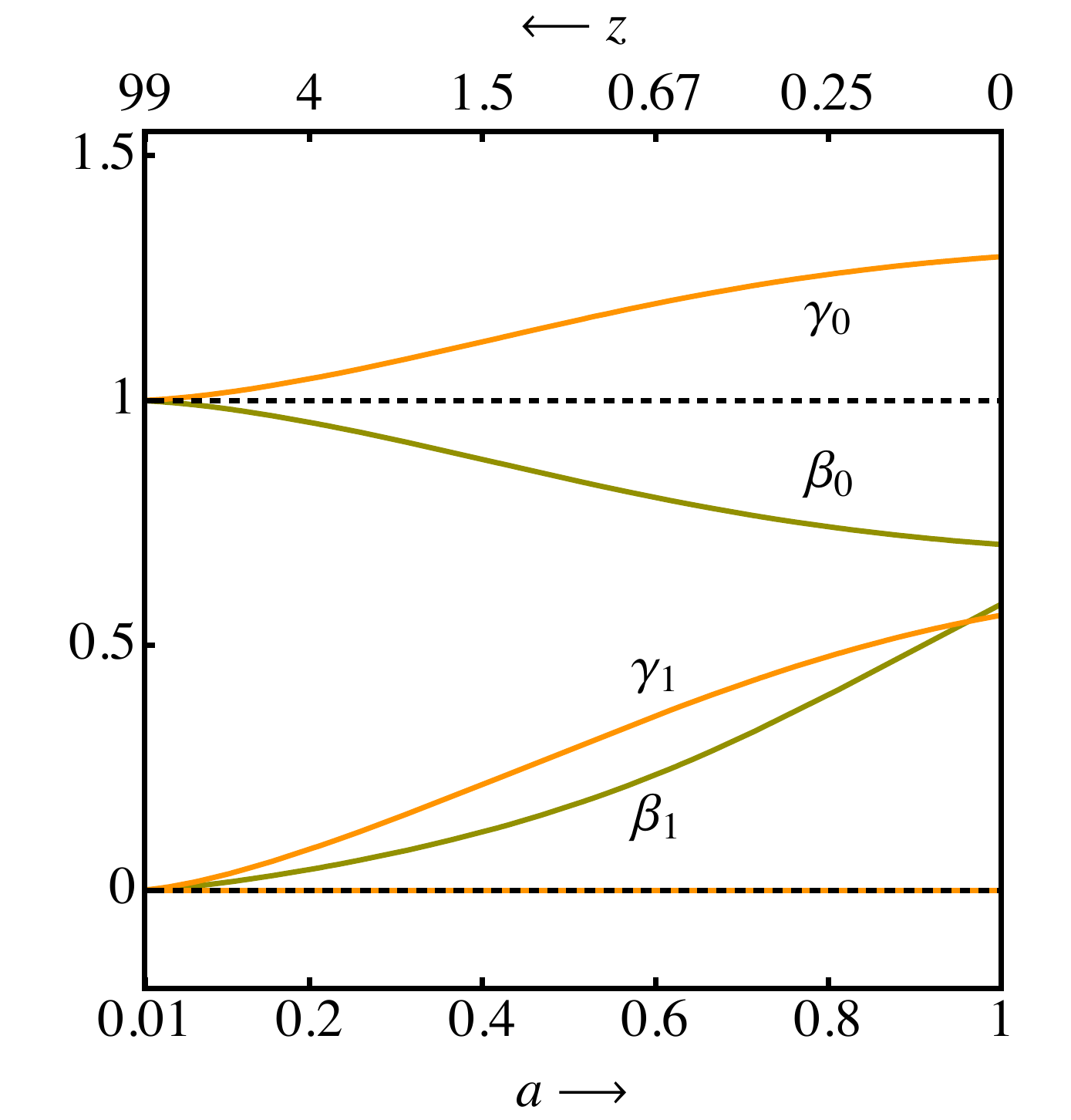} 
    \caption{The dimensionless coefficient functions characterizing the relationship between the metric perturbations and matter distribution are show above for $\Lambda$CDM(dashed lines) and DGP braneworld model (solid lines). The variation of $\beta_0$ with the scale factor in DGP can be interpreted as a variation of Newton's constant ``$G\beta_0$'' as far as growth of perturbations is concerned. Also note that for DGP, $\beta_0\ne\gamma_0$. In contrast to all the other examples considered, the coefficients of $aH/k$: $\beta_1,\gamma_1\ne0$. This is due to the junction conditions on the brane. The linear $aH/k$ term provides an intriguing signature of braneworld models. } 
        \label{fig: coeff_2}
 \end{figure}
A general way to understand the odd power in our $(aH/k)$ expansion is as follows. The Israel junction condition relates the first $(y)$ derivative of a metric perturbation normal to the brane at its surface to the 4D Einstein tensor and stress-energy tensor in the brane (see equations (\ref{Israel}),(\ref{GN}) and (\ref{RHS})). The 5D vacuum Einstein equations in the bulk provide homogeneous second-order linear differential equations for the metric perturbations. Just outside the brane, the operators in these equations will be dominated (in our large $k$ limit) by the 3-dimensional spatial derivatives (continuous across the brane surface), giving terms proportional to $k^2$. These must be balanced by terms proportional to $\partial_y^2$. Thus we see that $\partial_y$ must be proportional to $k$. 

We note that the above arguments are rather general. Although the $RHS$ of equation (\ref{Israel}) will be different in different braneworld models, due to the $LHS$ a linear $k^{-1}$ term will be present in most braneworld models. However, the coefficients might not be of $\mathcal{O}[1]$ as in the case of DGP.  \footnote{Another way of seeing the odd power of $aH/k$ in DGP case is through the form of a propagator which involves $\sqrt{\Box}$ \citep{Dvali:2000hr}.} The existence of this non-zero linear $aH/k$ term provides an exciting new signature for the DGP (or other ) braneworld models. Since on subhorizon scales $aH/k\ll1$, it is significantly easier to constrain the linear term compared to the quadratic one.
 
\begin{figure}
    \centering
    \includegraphics[width=2.7in]{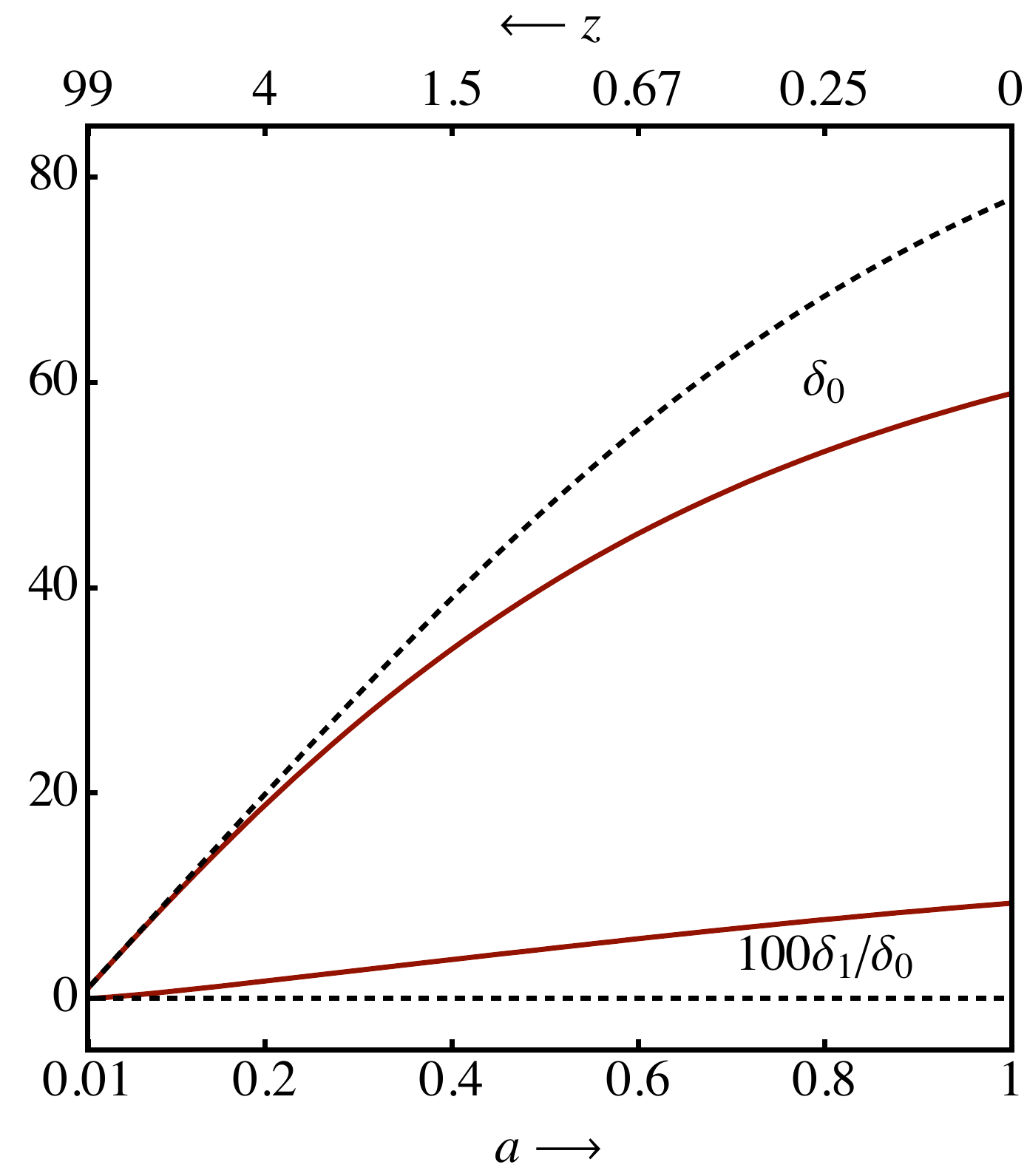} 
    \caption{The dimensionless coefficient functions characterizing growth of structure are show above for $\Lambda$CDM(dashed lines) and DGP braneworld model (solid lines). $\delta_0$ is the usual growth function on small scales, whereas $\delta_1$ characterizes the departures as we move to larger scales. In contrast to $\Lambda$CDM, for the DGP case $\delta_1\ne 0$. This could provide a distinct signature of braneworld models. }
        \label{fig: coeff_3}
 \end{figure}

We plot the coefficient functions in Figures 4 and 5. For DGP, we have assumed $\Omega_m=0.25$ and $\Omega_k=0$ for simplicity although this model is in tension with current data \citep{Song:2006jk}. The dotted lines represent the coefficients in $\Lambda$CDM. We have not calculated the coefficients $\beta_2,\gamma_2$ and $\delta_2$ of the $(aH/k)^2$ terms for the DGP case because they are expected to be subdominant compared to the linear $aH/k$ terms. 

We have ignored two important features in the DGP model, the strong coupling problem  and the ghost problem in the self accelerating branch (see for example \citealt{Deffayet:2001uk, Luty:2003vm}). The transition to the strong coupling regime happens at the Vainshtein radius $r_\star$ . For a localized matter distributions with Schwarzchild radius $r_g$, $r_\star\sim (r_g r_c^2)^{1/3}$. Using $r_c\sim 1/H_0$, for the largest localized distribution in our Universe, galaxy clusters ($M\sim 10^{14} M_{\odot}$), we get $r_\star < 10\, \rm{Mpc}$. This scale is well below the smallest scales where we intend to apply our framework. 
We are unaware of a calculation for $r_\star$, when considering distribution of matter on cosmological scales (which is not localized). To avoid the ghost problem one can choose the normal branch rather than the self-accelerating branch of the DGP model (see for example \citealt{Charmousis:2006pn}). A calculation similar to the one done in this section can be repeated for the normal branch, however in that case we do not have an accelerating universe. 
\end{subsection}

In this section we have calculated the coefficient functions for a few examples. Our aim was to give a flavor of the calculations rather than be exhaustive in the investigation of the models considered. It would be interesting to investigate these models in more detail in the context of these coefficient functions to see if there is come generic behavior across a large class of models. Based on the examples considered it might be tempting to conclude that $\beta_n-\gamma_n\ne0$ indicates physics beyond general relativity. However this is not so. For example a hypothetical dark energy component could also yield significant anisotropic stress. In the early Universe, a more standard source of anisotropic stress was provided by neutrinos. Nevertheless this difference could serve as an indicator of new physics in the matter or gravity sector. Another intriguing result was the presence of a term linear in $aH/k$ in the braneworld scenario, which could provide a unique signature of braneworld models. We have left out many possibilities including  Bekenstein's TeVeS \citep{Bekenstein:2004ne}, models with non-canonical kinetic terms \citep{Chiba:1999ka}, models of imperfect fluid dark energy with anisotropic stress \citep{Koivisto:2005mm}, and many others (see \citealt{Copeland:2006wr} for a review). We now turn our attention to observables and their relationship to the coefficient functions.

\end{section}
\begin{section}{Observational Implications}

We have outlined a procedure that allows many alternative, dynamical theories to GR with F$\Lambda$CDM cosmology to be explored within a common framework. Our approach has been devised with future observations in mind as its usefulness is limited to the observations that we expect will be the most prescriptive. We assume that the expansion history of the Universe is well constrained  through the distance redshift relation obtained from the apparent magnitude of Type 1a supernovae, the baryon acoustic oscillation scale and the ratio of  baryons to total matter in galaxy clusters. The large $k$ expansion connects the inhomogeneous nonrelativstic matter distribution to the perturbed metric in a universe of known (unperturbed) kinematical behavior, {\it i.e.} with a given relation $H(a)$ [or, equivalently, $a(t)$]. Our approach also presumes that the theories under consideration 
provide an understanding of how the distribution of observable entities such as galaxies relate to that of total mass. This allows us to focus on the manner in which structure can be observed to grow in the linear regime well within the horizon, which avoids the limitations imposed by cosmic variance considerations and the complications associated with gas dynamics. We further suppose that gravitational motion of baryonic matter and photons follows timelike and null geodesics  respectively in this spacetime. 

From an observational standpoint, our focus is on comoving length scales from $\sim 40$ Mpc to $\sim 400$ Mpc or equivalently $300\
\gtrsim l\gtrsim30$ at $z\sim1$, where we expect the effects to stand out the best. There are three types of observations that are likely to be relevant. Firstly, there are direct measurements of the two point correlation function and its evolution. Counting galaxies (or clusters) in three dimensions will lead to measurements of the evolution of the density function $\delta_m$ using future survey instruments such as LSST \citep{Tyson:2003kb, Zhan:2005rz} limited solely by cosmic variance as the photometric redshift accuracy and biasing errrors will be ignorable on these scales. We can construct the ratio of the matter power spectrum $P_{\delta_m}(k,a)$ at different redshifts to obtain constraints on $\delta_0(a),\delta_1(a)$ and $\delta_2(a)$. As discussed in the Introduction, by taking ratios, we can eliminate the need for knowing the initial conidtions $\delta_{mi}({\bf k})$. 

\begin{equation}
\begin{aligned}
\frac{P_{\delta_m}(k,a_2)}{P_{\delta_m}(k,a_1)}=\left[\frac{\delta_0(a_2)}{\delta_0(a_1)}\right]^2\left[1+\left\{\frac{aH}{k}\frac{\delta_0}{\delta_1}\right\}^{a_2}_{a_1}+\hdots\right]\nonumber
\end{aligned}
\end{equation}
where $\left\{f(k,a)\right\}^{a_2}_{a_1}\equiv f(k,a_2)-f(k,a_1)$.

The second type of observation that will be carried out involves departures from the Hubble flow. These are dominated by the potential function $\Phi$. Under our assumptions, galaxies will follow timelike geodesics and satisfy the linear conservation equations relating their peculiar velocities to $\Phi$. 

Finally there are weak lensing observations which depend upon the sum, $\Phi+\Psi$, presuming photons follow null geodesics.  These then allow us to track the evolution of $\Psi$. A combination of these measurements would not only allow us to understand the scale dependent evolution of $\Phi,\Psi$ and $\delta_m$ but also allow us to probe the relationship between them. For example, using our ansatz, one can obtain constraints on the coefficient functions by comparing the correlation functions for the potentials, $P_{\Phi+\Psi}(k,a)$ (provided by lensing tomography) and the nonrelativistic matter overdensity $P_{\delta_m}(k,a)$ (provided by growth of structure measurements) using
$$k^4P_{\Phi+\Psi}\propto P_{\delta_m}(\beta_0+\gamma_0)^2\left[1+2\left(\frac{\beta_1+\gamma_1}{\beta_0+\gamma_0}\right)\left(\frac{aH}{k}\right)+...\right]$$
Comparing the matter and potential power spectrum allows us to constrain the coefficient function without worrying about the initial conditions, though one would still have to obtain this ratio at different redshifts to constrain the time evolution of the coefficient functions. 

In this exploratory paper, we have discussed only a handful of observations that can allow is to constrain the coefficient functions. In addition to the observations mentioned above, we list a few other observations that we think might be relevant for our framework. The matter and potential fluctuations at the last scattering surface can be compared to their counterparts at late times, as long as we restrict ourselves to linear subhorizon scales. The same is true for BAO measurements (see discussion of range of scales at the end of Section 2). Recently, a $3\sigma$ detection of lensing of the CMB at large $l$, has been reported by the ACBAR group \citep{Reichardt:2008ay}. This measurement probes the distribution and evolution of potentials after last scattering, and can also be used for constraining the coefficient functions \citep{Calabrese:2008rt}. With the Planck mission \citep{Planck:2006uk}, such constraints are expected to improve significantly. Another exciting probe of the three dimensional matter distribution may be provided by the 21 cm observations (see for example \citealt{Mao:2008ug} and references therein).

We have limited ourselves to the linear regime. On small scales, the nonlinear matter power spectrum and its evolution can play a role in the observations discussed above. The linear to nonlinear mapping discussed in \citep{Smith:2002dz} can be used for this purpose. However, without understanding the theories under consideration in the nonlinear regime, this is not fully robust. 
 
Recall that $\{\beta_n,\gamma_n,\delta_n\}$ with ($n=0,1,2$) are functions of the scale factor, $a$. If the observations are to be done in a limited range of redshifts  then Taylor expanding the coefficient functions around the central value of the redshift might be a  simple and model independent way of characterizing these coefficient functions in terms of a few parameters. From a theoretical perspective, the coefficient functions will depend on relevant parameters in the theory or model under consideration.  A detailed investigation of the parameterization of the coefficient functions and the possible constraints that can be obtained from current and future observations is beyond the scope of this paper. For a more detailed discussion of the observations for distinguishing different models of modified gravity and dark energy we refer the reader to some of the references cited at the end of  Section 1 in this paper. 

\end{section}
\begin{section}{Discussion}

We have outlined a procedure that can be used to test the application of general relativity (more specifically F$\Lambda$CDM) on cosmological scales in the context where it is most likely to fail and in the regime where observations should be most sensitive to measuring a departure from the general relativistic prediction. The scales are large enough to avoid the complications from nonlinearities and gas physics, yet small enough to avoid strong limitations to the interpretation of observations posed by cosmic variance. 

Our procedure assumes that (i) The geometry and kinematics of the Universe is understood (ii) baryons and photons behave as ideal test particles following geodesics of the cosmological metric. Given these assumptions, at late times, it is the relationship between the cosmological metric and the nonrelativistic matter distribution (along with their respective evolution) that provides a test for alternatives GR with a cosmological constant and cold dark matter.  To probe the dynamics of gravity (or any additional fields) we provided an  ansatz, equation(\ref{ansatz}), which gave a relationship between the cosmological metric and nonrelativistic matter perturbations in the linear, subhorizon regime. This form of the ansatz is consistent with a large class of theories with the differences between different theories evident in the coefficient functions $\{\beta_n(a),\gamma_n(a),\delta_n(a)\}$ with $n=0,1, 2$.  It is hoped that three scalar functions, the nonrelativistic matter overdensity $\delta_m$ and the metric potentials $\Phi$ and $\Psi$ can be measured over the next decade, providing constraints on the coefficient functions. Constraining these coefficient functions provides observers with concrete targets for testing gravity in a scale dependent manner. 

Our goal was to provide a perturbative framework, similar in spirit to the PPN formalism for testing gravity on solar system scales. However unlike the PPN case, we were left with coefficient functions that depend on the scale factor rather than constant coefficients. Although we have not done so in this paper, if the observations are limited to a small range of scale factors, it is possible to characterize these coefficient functions using a few parameters by expanding around a given scale factor at which the observations are centered. 

With our choice of scales, we have restricted ourselves to linear, subhorizon evolution. We leave the connection between superhorizon and subhorizon evolution as well as consideration of nonlinearities for the future. Although, we have restricted ourselves to scalar perturbations, the framework could be extended to include vector and tensor perturbations. 

\end{section}
\begin{section}{Acknowledgments}
We thank the members of KIPAC at Stanford University and the Stanford Gravity Probe-B theory group for helpful discussions. We would also like to thank an anonymous referee for a number of insightful comments and suggestions. MAA would like to acknowledge the financial support of a Stanford Graduate Fellowship.  RDB acknowledges support from the National Science Foundation grant AST05-07732. This work was supported in part by the U.S. Department of Energy under contract number DE-AC02-76SF00515.
\end{section}
\def \aap {A\&A} 
\def \statisci {Statis. Sci.}
\def \pre {Phys.\ Rev.\ E}
\def \apj {ApJ}
\def \apjl {ApJL}
\def \apjs {ApJS}
\def \mnras {MNRAS}
\def \prd {Phys.\ Rev.\ D}
\def \prl {Phys.\ Rev.\ Lett.}

\bibliography{LargeKexp3}	

\begin{thebibliography}{}

\bibitem[\protect\citeauthoryear{{Allen}, {Rapetti}, {Schmidt}, {Ebeling},
  {Morris} \& {Fabian}}{{Allen} et~al.}{2008}]{2008MNRAS.383..879A}
{Allen} S.~W.,  {Rapetti} D.~A.,  {Schmidt} R.~W.,  {Ebeling} H.,  {Morris}
  R.~G.,    {Fabian} A.~C.,  2008, \mnras, 383, 879

\bibitem[\protect\citeauthoryear{Allen, Schmidt \& Fabian}{Allen
  et~al.}{2002}]{Allen:2002sr}
Allen S.~W.,  Schmidt R.~W.,    Fabian A.~C.,  2002, \mnras, 334, L11

\bibitem[\protect\citeauthoryear{{Amendola}, {Kunz} \& {Sapone}}{{Amendola}
  et~al.}{2008}]{2008JCAP...04..013A}
{Amendola} L.,  {Kunz} M.,    {Sapone} D.,  2008, Journal of Cosmology and
  Astro-Particle Physics, 4, 13

\bibitem[\protect\citeauthoryear{Armendariz-Picon}{Armendariz-Picon}{2004}]{Ar%
mendarizPicon:2004pm}
Armendariz-Picon C.,  2004, JCAP, 0407, 007

\bibitem[\protect\citeauthoryear{Armendariz-Picon, Mukhanov \&
  Steinhardt}{Armendariz-Picon et~al.}{2000}]{ArmendarizPicon:2000dh}
Armendariz-Picon C.,  Mukhanov V.~F.,    Steinhardt P.~J.,  2000, Phys. Rev.
  Lett., 85, 4438

\bibitem[\protect\citeauthoryear{Astier et~al.,}{Astier
  et~al.}{2006}]{Astier:2005qq}
Astier P.,  et~al., 2006, Astron. Astrophys., 447, 31

\bibitem[\protect\citeauthoryear{Bardeen}{Bardeen}{1980}]{Bardeen:1980kt}
Bardeen J.~M.,  1980, Phys. Rev., D22, 1882

\bibitem[\protect\citeauthoryear{Bashinsky}{Bashinsky}{2007}]{Bashinsky:2007yc}
Bashinsky S.,  2007, arXiv:0707.0692 [astro-ph]

\bibitem[\protect\citeauthoryear{Bean}{Bean}{2001}]{Bean:2001ys}
Bean R.,  2001, Phys. Rev., D64, 123516

\bibitem[\protect\citeauthoryear{Bean, Bernat, Pogosian, Silvestri \&
  Trodden}{Bean et~al.}{2007}]{Bean:2006up}
Bean R.,  Bernat D.,  Pogosian L.,  Silvestri A.,    Trodden M.,  2007, Phys.
  Rev., D75, 064020

\bibitem[\protect\citeauthoryear{Bekenstein}{Bekenstein}{2004}]{Bekenstein:200%
4ne}
Bekenstein J.~D.,  2004, Phys. Rev., D70, 083509

\bibitem[\protect\citeauthoryear{Bertschinger}{Bertschinger}{2006}]{Bertsching%
er:2006aw}
Bertschinger E.,  2006, Astrophys. J., 648, 797

\bibitem[\protect\citeauthoryear{Bertschinger \& Zukin}{Bertschinger \&
  Zukin}{2008}]{Bertschinger:2008zb}
Bertschinger E.,  Zukin P.,  2008, arXiv:0801.2431 [astro-ph]

\bibitem[\protect\citeauthoryear{{Bludman}}{{Bludman}}{2007}]{2007soch.conf...%
.9B}
{Bludman} S.,  2007, in VI Reunion Anual Sociedad Chilena de Astronomia
  (SOCHIAS) {Cosmological Acceleration: Dark Energy or Modified Gravity?}.
pp~9--+

\bibitem[\protect\citeauthoryear{Boisseau, Esposito-Farese, Polarski \&
  Starobinsky}{Boisseau et~al.}{2000}]{Boisseau:2000pr}
Boisseau B.,  Esposito-Farese G.,  Polarski D.,    Starobinsky A.~A.,  2000,
  Phys. Rev. Lett., 85, 2236

\bibitem[\protect\citeauthoryear{Calabrese, Slosar, Melchiorri, Smoot \&
  Zahn}{Calabrese et~al.}{2008}]{Calabrese:2008rt}
Calabrese E.,  Slosar A.,  Melchiorri A.,  Smoot G.~F.,    Zahn O.,  2008,
  arXiv:0803.2309 [astro-ph]

\bibitem[\protect\citeauthoryear{Caldwell, Cooray \& Melchiorri}{Caldwell
  et~al.}{2007}]{Caldwell:2007cw}
Caldwell R.,  Cooray A.,    Melchiorri A.,  2007, astro-ph/0703375

\bibitem[\protect\citeauthoryear{Carroll}{Carroll}{2001}]{Carroll:2000fy}
Carroll S.~M.,  2001, Living Rev. Rel., 4, 1

\bibitem[\protect\citeauthoryear{Carroll, Duvvuri, Trodden \& Turner}{Carroll
  et~al.}{2004}]{Carroll:2003wy}
Carroll S.~M.,  Duvvuri V.,  Trodden M.,    Turner M.~S.,  2004, Phys. Rev.,
  D70, 043528

\bibitem[\protect\citeauthoryear{Charmousis, Gregory, Kaloper \&
  Padilla}{Charmousis et~al.}{2006}]{Charmousis:2006pn}
Charmousis C.,  Gregory R.,  Kaloper N.,    Padilla A.,  2006, JHEP, 10, 066

\bibitem[\protect\citeauthoryear{Chiba, Okabe \& Yamaguchi}{Chiba
  et~al.}{2000}]{Chiba:1999ka}
Chiba T.,  Okabe T.,    Yamaguchi M.,  2000, Phys. Rev., D62, 023511

\bibitem[\protect\citeauthoryear{Chiba, Smith \& Erickcek}{Chiba
  et~al.}{2007}]{Chiba:2006jp}
Chiba T.,  Smith T.~L.,    Erickcek A.~L.,  2007, Phys. Rev., D75, 124014

\bibitem[\protect\citeauthoryear{Cole et~al.,}{Cole
  et~al.}{2005}]{Cole:2005sx}
Cole S.,  et~al., 2005, \mnras, 362, 505

\bibitem[\protect\citeauthoryear{Copeland, Sami \& Tsujikawa}{Copeland
  et~al.}{2006}]{Copeland:2006wr}
Copeland E.~J.,  Sami M.,    Tsujikawa S.,  2006, Int. J. Mod. Phys., D15, 1753

\bibitem[\protect\citeauthoryear{Deffayet}{Deffayet}{2002}]{Deffayet:2002fn}
Deffayet C.,  2002, Phys. Rev., D66, 103504

\bibitem[\protect\citeauthoryear{Deffayet, Dvali, Gabadadze \&
  Vainshtein}{Deffayet et~al.}{2002}]{Deffayet:2001uk}
Deffayet C.,  Dvali G.~R.,  Gabadadze G.,    Vainshtein A.~I.,  2002, Phys.
  Rev., D65, 044026

\bibitem[\protect\citeauthoryear{Dvali, Gabadadze \& Porrati}{Dvali
  et~al.}{2000}]{Dvali:2000hr}
Dvali G.~R.,  Gabadadze G.,    Porrati M.,  2000, Phys. Lett., B485, 208

\bibitem[\protect\citeauthoryear{Eisenstein et~al.,}{Eisenstein
  et~al.}{2005}]{Eisenstein:2005su}
Eisenstein D.~J.,  et~al., 2005, Astrophys. J., 633, 560

\bibitem[\protect\citeauthoryear{Farrar \& Peebles}{Farrar \&
  Peebles}{2004}]{Farrar:2003uw}
Farrar G.~R.,  Peebles P. J.~E.,  2004, Astrophys. J., 604, 1

\bibitem[\protect\citeauthoryear{{Faulkner}, {Tegmark}, {Bunn} \&
  {Mao}}{{Faulkner} et~al.}{2007}]{2007PhRvD..76f3505F}
{Faulkner} T.,  {Tegmark} M.,  {Bunn} E.~F.,    {Mao} Y.,  2007, \prd, 76,
  063505

\bibitem[\protect\citeauthoryear{Freedman et~al.,}{Freedman
  et~al.}{2001}]{Freedman:2000cf}
Freedman W.~L.,  et~al., 2001, Astrophys. J., 553, 47

\bibitem[\protect\citeauthoryear{Frolov}{Frolov}{2008}]{Frolov:2008uf}
Frolov A.~V.,  2008, arXiv:0803.2500 [astro-ph]

\bibitem[\protect\citeauthoryear{{Hu}}{{Hu}}{2008}]{2008arXiv0801.2433H}
{Hu} W.,  2008, ArXiv e-prints, 801

\bibitem[\protect\citeauthoryear{Hu \& Eisenstein}{Hu \&
  Eisenstein}{1999}]{Hu:1998tj}
Hu W.,  Eisenstein D.~J.,  1999, Phys. Rev., D59, 083509

\bibitem[\protect\citeauthoryear{{Hu} \& {Sawicki}}{{Hu} \&
  {Sawicki}}{2007}]{2007PhRvD..76j4043H}
{Hu} W.,  {Sawicki} I.,  2007, \prd, 76, 104043

\bibitem[\protect\citeauthoryear{{Hui} \& {Parfrey}}{{Hui} \&
  {Parfrey}}{2007}]{2007arXiv0712.1162H}
{Hui} L.,  {Parfrey} K.~P.,  2007, ArXiv e-prints, 712

\bibitem[\protect\citeauthoryear{Huterer \& Linder}{Huterer \&
  Linder}{2007}]{Huterer:2006mv}
Huterer D.,  Linder E.~V.,  2007, Phys. Rev., D75, 023519

\bibitem[\protect\citeauthoryear{Ishak, Upadhye \& Spergel}{Ishak
  et~al.}{2006}]{Ishak:2005zs}
Ishak M.,  Upadhye A.,    Spergel D.~N.,  2006, Phys. Rev., D74, 043513

\bibitem[\protect\citeauthoryear{Jain \& Zhang}{Jain \&
  Zhang}{2007}]{Jain:2007yk}
Jain B.,  Zhang P.,  2007, arXiv:0709.2375 [astro-ph]

\bibitem[\protect\citeauthoryear{Koivisto \& Mota}{Koivisto \&
  Mota}{2006}]{Koivisto:2005mm}
Koivisto T.,  Mota D.~F.,  2006, Phys. Rev., D73, 083502

\bibitem[\protect\citeauthoryear{Komatsu et~al.,}{Komatsu
  et~al.}{2008}]{Komatsu:2008hk}
Komatsu E.,  et~al., 2008, arXiv:0803.0547 [astro-ph]

\bibitem[\protect\citeauthoryear{{Linder} \& {Cahn}}{{Linder} \&
  {Cahn}}{2007}]{2007APh....28..481L}
{Linder} E.~V.,  {Cahn} R.~N.,  2007, Astroparticle Physics, 28, 481

\bibitem[\protect\citeauthoryear{Lue, Scoccimarro \& Starkman}{Lue
  et~al.}{2004}]{Lue:2003ky}
Lue A.,  Scoccimarro R.,    Starkman G.,  2004, Phys. Rev., D69, 044005

\bibitem[\protect\citeauthoryear{Luty, Porrati \& Rattazzi}{Luty
  et~al.}{2003}]{Luty:2003vm}
Luty M.~A.,  Porrati M.,    Rattazzi R.,  2003, JHEP, 09, 029

\bibitem[\protect\citeauthoryear{{Mantz}, {Allen}, {Ebeling} \&
  {Rapetti}}{{Mantz} et~al.}{2008}]{2008MNRAS.387.1179M}
{Mantz} A.,  {Allen} S.~W.,  {Ebeling} H.,    {Rapetti} D.,  2008, \mnras, 387,
  1179

\bibitem[\protect\citeauthoryear{Mao, Tegmark, McQuinn, Zaldarriaga \&
  Zahn}{Mao et~al.}{2008}]{Mao:2008ug}
Mao Y.,  Tegmark M.,  McQuinn M.,  Zaldarriaga M.,    Zahn O.,  2008,
  arXiv:0802.1710 [astro-ph]

\bibitem[\protect\citeauthoryear{Nojiri \& Odintsov}{Nojiri \&
  Odintsov}{2007}]{Nojiri:2006ri}
Nojiri S.,  Odintsov S.~D.,  2007, Int. J. Geom. Meth. Mod. Phys., 4, 115

\bibitem[\protect\citeauthoryear{Percival et~al.,}{Percival
  et~al.}{2007}]{Percival:2007yw}
Percival W.~J.,  et~al., 2007, \mnras, 381, 1053

\bibitem[\protect\citeauthoryear{Perlmutter et~al.,}{Perlmutter
  et~al.}{1999}]{Perlmutter:1998np}
Perlmutter S.,  et~al., 1999, Astrophys. J., 517, 565

\bibitem[\protect\citeauthoryear{Perrotta, Baccigalupi \& Matarrese}{Perrotta
  et~al.}{2000}]{Perrotta:1999am}
Perrotta F.,  Baccigalupi C.,    Matarrese S.,  2000, Phys. Rev., D61, 023507

\bibitem[\protect\citeauthoryear{Planck-Collaboration}{Planck-Collaboration}{2%
006}]{Planck:2006uk}
Planck-Collaboration 2006, astro-ph/0604069

\bibitem[\protect\citeauthoryear{Pogosian \& Silvestri}{Pogosian \&
  Silvestri}{2008}]{Pogosian:2007sw}
Pogosian L.,  Silvestri A.,  2008, Phys. Rev., D77, 023503

\bibitem[\protect\citeauthoryear{Rapetti, Allen, Amin \& Blandford}{Rapetti
  et~al.}{2007}]{Rapetti:2006fv}
Rapetti D.,  Allen S.~W.,  Amin M.~A.,    Blandford R.~D.,  2007, \mnras, 375,
  1510

\bibitem[\protect\citeauthoryear{Ratra \& Peebles}{Ratra \&
  Peebles}{1988}]{Ratra:1987rm}
Ratra B.,  Peebles P. J.~E.,  1988, Phys. Rev., D37, 3406

\bibitem[\protect\citeauthoryear{Reichardt et~al.,}{Reichardt
  et~al.}{2008}]{Reichardt:2008ay}
Reichardt C.~L.,  et~al., 2008, arXiv:0801.1491 [astro-ph]

\bibitem[\protect\citeauthoryear{Riess et~al.,}{Riess
  et~al.}{1998}]{Riess:1998cb}
Riess A.~G.,  et~al., 1998, Astron. J., 116, 1009

\bibitem[\protect\citeauthoryear{Riess et~al.,}{Riess
  et~al.}{2004}]{Riess:2004nr}
Riess A.~G.,  et~al., 2004, Astrophys. J., 607, 665

\bibitem[\protect\citeauthoryear{Santiago, Kalligas \& Wagoner}{Santiago
  et~al.}{1998}]{Santiago:1998ae}
Santiago D.~I.,  Kalligas D.,    Wagoner R.~V.,  1998, Phys. Rev., D58, 124005

\bibitem[\protect\citeauthoryear{Schimd, Uzan \& Riazuelo}{Schimd
  et~al.}{2005}]{Schimd:2004nq}
Schimd C.,  Uzan J.-P.,    Riazuelo A.,  2005, Phys. Rev., D71, 083512

\bibitem[\protect\citeauthoryear{Smith et~al.,}{Smith
  et~al.}{2003}]{Smith:2002dz}
Smith R.~E.,  et~al., 2003, \mnras, 341, 1311

\bibitem[\protect\citeauthoryear{Song, Hu \& Sawicki}{Song
  et~al.}{2007}]{Song:2006ej}
Song Y.-S.,  Hu W.,    Sawicki I.,  2007, Phys. Rev., D75, 044004

\bibitem[\protect\citeauthoryear{Song, Sawicki \& Hu}{Song
  et~al.}{2007}]{Song:2006jk}
Song Y.-S.,  Sawicki I.,    Hu W.,  2007, Phys. Rev., D75, 064003

\bibitem[\protect\citeauthoryear{Spergel et~al.,}{Spergel
  et~al.}{2007}]{Spergel:2006hy}
Spergel D.~N.,  et~al., 2007, Astrophys. J. Suppl., 170, 377

\bibitem[\protect\citeauthoryear{Tegmark et~al.,}{Tegmark
  et~al.}{2004}]{Tegmark:2003ud}
Tegmark M.,  et~al., 2004, Phys. Rev., D69, 103501

\bibitem[\protect\citeauthoryear{{Tsujikawa}}{{Tsujikawa}}{2007}]{2007PhRvD..7%
6b3514T}
{Tsujikawa} S.,  2007, \prd, 76, 023514

\bibitem[\protect\citeauthoryear{Tyson}{Tyson}{2002}]{Tyson:2003kb}
Tyson J.~A.,  2002, Proc. SPIE Int. Soc. Opt. Eng., 4836, 10

\bibitem[\protect\citeauthoryear{Will}{Will}{2006}]{lrr-2001-4}
Will C.~M.,  2006, Living Reviews in Relativity, 9

\bibitem[\protect\citeauthoryear{Zhan, Knox, Tyson \& Margoniner}{Zhan
  et~al.}{2006}]{Zhan:2005rz}
Zhan H.,  Knox L.,  Tyson A.,    Margoniner V.,  2006, Astrophys. J., 640, 8

\bibitem[\protect\citeauthoryear{{Zhang}, {Liguori}, {Bean} \&
  {Dodelson}}{{Zhang} et~al.}{2007}]{2007PhRvL..99n1302Z}
{Zhang} P.,  {Liguori} M.,  {Bean} R.,    {Dodelson} S.,  2007, Physical Review
  Letters, 99, 141302

\end{thebibliography}
\bibliographystyle{mn2e}
\label{142}
\end{document}